\documentclass[12pt,cite,epsf,epsfig,psfrag]{article}
\usepackage{epsfig}
\usepackage{amsmath, graphics, setspace}
\usepackage{graphicx}
\usepackage{amssymb}
\usepackage{subfigure}

\def\be{\begin{equation}}
\def\ee{\end{equation}}
\def\bea{\begin{eqnarray}}
\def\eea{\end{eqnarray}}

\newcommand{\beqal}{\begin{eqnarray}\label}
\newcommand{\beqa}{\begin{eqnarray}}
\newcommand{\eeqa}{\end{eqnarray}}

\newcommand{\mathsym}[1]{{}}
\newcommand{\unicode}[1]{{}}

\begin{document}

\begin{titlepage}
\begin{center}

\vskip .2in

{\Large \bf Back reaction effects on the dynamics of heavy probes in heavy quark cloud}
\vskip .5in

{\bf Shankhadeep Chakrabortty}$^{a}$\footnote{s.chakrabortty@rug.nl} and {\bf Tanay K. Dey} $^{b}$\footnote{tanay.dey@gmail.com} \\
\vskip .1in
{ $^{a}$ Van Swinderen Institute for Particle Physics and Gravity, University of Groningen,
\\ Nijenborgh 4, 9747 AG Groningen, The Netherlands.\\
\vskip .3in

 $^{b}$Department of Physics, \\
Sikkim Manipal Institute of Technology,\\
Majitar, Rongpo, East Sikkim, Sikkim-737136, INDIA\\}
\end{center}

\begin{center} {\bf ABSTRACT}

\end{center}
\begin{quotation}\noindent
\baselineskip 15pt

We holographically study the effect of back reaction on the hydrodynamical properties of $\mathcal{N} = 4$ strongly coupled super Yang-Mills (SYM) thermal plasma. The back reaction we consider arises from the presence of static heavy quarks uniformly distributed over $\mathcal{N} = 4$ SYM plasma. In order to study the hydrodynamical properties, we use heavy quark as well as heavy quark-antiquark bound state as probes and compute the jet quenching parameter, screening length and binding energy. We also consider the rotational dynamics of  heavy probe quark in the back-reacted plasma and analyse associated energy loss. We observe that the presence of back reaction enhances the energy-loss in the thermal plasma. Finally, we show that there is no effect of angular drag on the rotational motion of quark-antiquark bound state probing the back reacted thermal plasma.

\end{quotation}
\end{titlepage}
\vfill
\eject

\section{Introduction}\label{intro}
The recent experimental results obtained at the Relativistic Heavy Ion Collider (RHIC)
and the Large Hadron Collider (LHC) indicate that a deconfined plasma phase consisted of free quarks and gluons (QGP) has been created at high temperature and high number density 
\cite{Adcox:2004mh, Adams:2005dq, Back:2004je, Shuryak:2008eq, Shuryak:2004cy}.
Further, the interaction between the high energetic parton probes and the QGP medium signifies
that the associated free quarks and gluons are strongly coupled \cite{Baier:1996kr, Eskola:2004cr}.
From the theoretical point of view, among the pre-existing successful theories of quantum chromodynamics,
the perturbative QCD and the lattice methods turn out to be inadequate to address the strong coupling issues.
On the other hand, the $gauge/gravity$ correspondence seems to be a promising theoretical candidate  since it has been widely utilized to study a large class of previously inaccessible strongly coupled
gauge theories \cite{Maldacena:1997re, Witten:1998qj, Gubser:1998bc, Aharony:1999ti}.
However, to make use of this correspondence we need to know the exact gravity dual of real QCD at strong coupling
and that is not well-understood till date. Nevertheless, the $gauge/gravity$ correspondence can extract some
universal properties of a large class of strongly coupled theories having well-defined gravity duals. Interestingly, those universal properties qualitatively agree with the experimental data associated with strong coupling phase of QGP \cite{Policastro:2001yc, Kovtun:2003wp, Buchel:2003tz, Teaney:2003kp, Chakrabarti:2010xy}.
 Moreover, the correspondence holds true for some
strongly coupled gauge theories exhibiting some QCD like features such as chiral
 symmetry breaking, confinement to deconfinement crossover etc \cite{Sakai:2004cn,
Gursoy:2007cb, Cai:2012xh}.

Along this line of development, within the regime of gauge/gravity correspondence, there has been a number of seminal works to
 obtain a better theoretical understanding of strongly coupled QGP phase. For example, the dissipative dynamics of an external heavy quark probing
through the $\mathcal{N} =4$ SYM plasma is holographically computed in
\cite{Herzog:2006gh, Gubser:2006bz}.
The rate of radiative energy loss of an external quark rotating in the $\mathcal{N} =4$ SYM plasma
is successfully addressed in \cite{Fadafan:2008bq}. Furthermore, the holographic technique
to compute the jet quenching parameter carrying a measure of suppression of
the heavy quark spectrum with high transverse momentum due
to the medium induced scattering has been first prescribed in \cite{Liu:2006ug}. The non-perturbative dynamics of heavy probe mesons moving through the
$\mathcal{N} =4$ SYM plasma has been studied and the corresponding quark-antiquark binding energy as well as screening length are
qualitatively estimated in \cite{Liu:2006nn}.
The holographic understanding of the Brownian motion of an external probe quark is achieved in \cite{deBoer:2008gu, Son:2009vu}.
There has been a lot of further generalisations along this direction of research \cite{Caceres:2006dj, Matsuo:2006ws, Chernicoff:2012iq, Chernicoff:2008sa, Herzog:2007kh,
VazquezPoritz:2008nw, Roy:2009sw, Chakrabortty:2011sp, Cai:2012eh, CasalderreySolana:2006rq, Gubser:2006nz,
CasalderreySolana:2007qw, Giataganas:2012zy, Atmaja:2012jg, Chernicoff:2012gu, Chernicoff:2012bu, Guijosa:2011hf, Fadafan:2012qu,
Chakraborty:2012dt, Fischler:2012ff, Gursoy:2010aa, Banerjee:2013rca, Giecold:2009cg,
Chakrabortty:2013kra, Chakrabortty:2014kma}.

In spite of several such developments, except in the very
few examples \cite{Bigazzi:2011it, Kumar:2012ui, Nunez:2010sf}, it remains very difficult to study the strongly coupled boundary gauge theory with large number of flavour quarks. The introduction of the flavour quarks in the boundary theory corresponds to adding an extra stack of $N_f$ flavour branes probing the pre-existed $N_c$ number of colour branes in the dual gravity \cite{Karch:2002sh}. The addition of these flavour branes exerts a back reaction of the order of $\frac{N_f}{N_c}$ on the bulk geometry. Therefore, the back reaction can not be neglected in the presence of large number of flavour branes ($ N_f \sim N_c^2 $ or more) even in the large $N_c$ limit.  The difficulty of going beyond the probe approximation motivated one of us to construct a $back reacted$ gravity background without $\textit{any}$ approximation \cite{Chakrabortty:2011sp}. The gravity background is realised as an AdS black hole back reacted in the presence of a uniform distribution of large number of fundamental strings. These strings are
assumed to be non-interacting, static and infinitely long. One of the end points of each string is attached to the boundary and the body of the string is aligned along the radial direction. The bulk space time gets deformed due to the back reaction of the string distribution. The back reacted
geometry is explicitly computable by solving Einstein equation of motion with negative cosmological constant sourced by the uniform
string distribution. It turns out to be a deformed black hole in AdS space time parameterized by the mass
and density of the strings. In five dimensional space time the solution reads as,
\begin{equation}\label{dualgrav}
ds^2= f(u)[-h(u)dt^2 + dx^2 + dy^2 +  dz^2 + {du^2\over h(u)}],
\end{equation}
where
\begin{equation}
f(u)= {l^2\over u^2} \quad\quad{\rm and}\quad\quad h(u)= 1-{2m u^4\over l^6} - {2\over 3}{b u^3\over l^4}.\nonumber
\end{equation}
Here, $b$ is the string cloud density, $u$ is the radial coordinate of AdS space with boundary at $u=0$ and $l$ is the radius of AdS space.
The radius of horizon can be constructed by solving the equation,
\begin{eqnarray}
 h(u_{+})= 1-{2m u_{+}^4\over l^6} - {2\over 3}{b u_{+}^3\over l^4}=0.
\label{hor}
\end{eqnarray}
The black hole geometry (\ref{dualgrav}) turns out to be stable under vector and tensor perturbation. 

The back reacted geometry is holographically dual to a system of large number of heavy, static flavour quarks uniformly distributed over the $\mathcal{N} = 4$ $SU(N_c)$ SYM thermal plasma. It is important to note that in the boundary theory, the SYM plasma together with the quark distribution is effectively considered as back reacted plasma. Using the holographic method applicable to the dual black hole background, dissipative force imparted by the back reacted thermal plasma on an external heavy probe quark has been studied \cite{Chakrabortty:2011sp}.\footnote{In \cite{Kumar:2012ui},  the author has considered a back reaction on a ten dimensional type IIB super gravity background
due to the presence of a uniform
distribution of strings preserving translational and SO(6) rotational symmetry and interacting with the Neveu-Schwarz two-form flux. Furthermore, by performing a dimensional reduction over this ten dimensional back reacted geometry an effective five dimensional geometry has been constructed and a certain asymptotic limit of this effective five dimensional background reproduces the same  asymptotically locally AdS geometry previously mentioned in (\ref{dualgrav}).}

In continuation of the earlier work, in this paper we aim to study the effect of back reaction on the jet quenching parameter $\hat{q}$ ,
screening length ($L_s$) and binding energy of a quark-antiquark pair ($q\bar{q}$).
We also analyse the rotational dynamics of an external heavy probe quark as well as heavy probe $q\bar{q}$ bound state in the back reacted plasma.

To elaborate further, following the holographic prescription mentioned in \cite{Liu:2006ug},
we compute a phenomenological transport coefficient,
namely jet quenching parameter($\hat{q}$) and study the effect of back reaction on this parameter.
In the holographic computation, the quark-antiquark
pair is mapped into the two endpoints of a fundamental string both of which
are attached to the boundary of the relevant dual background. The body of the string hangs down along radial
coordinate of the bulk geometry.
Motivated by eikonal approximation \cite{Kovner:2003zj, Kovner:2001vi}, the holographic working
formula to calculate the jet quenching parameter is constructed by considering the correspondence between thermal expectation value
of the light-like Wilson loop operator in fundamental representation,
$\langle\mathcal{W}^F_{light-like}\rangle$ and the
exponential of the string world-sheet action S, $e^{iS}$. Similarly, following \cite{Liu:2006nn} we compute the binding energy and the
screening length ($L_s$) between a $q\bar{q}$ pair moving with a constant linear speed in the hot back reacted plasma.
The screening length is defined as the maximum separation between a $q\bar{q}$ pair beyond which the pair breaks off and gets screened in
the thermal medium. Here, the holographic dual to the $q\bar{q}$ pair is similar to the one considered in the context of jet quenching parameter.
The study of $L_s$ and the binding energy requires the correspondence between thermal expectation
value of the time like Wilson loop $\langle \mathcal{W}^{fund}(\mathcal{C}_{time-like})\rangle$
traced out by a $q\bar{q}$ pair and $e^{iS}$, where $S$ is the string world sheet action.
 Consequently, we aim to obtain the $L_s$ from the boundary condition on radial coordinate of the background geometry and
 discuss how the back reaction modifies the original computation given in \cite{Liu:2006nn}.

In this paper we also investigate the effect of back reaction on the energy loss experienced by an external heavy probe quark rotating
along a circle of radius $\mathcal{R}$ with a constant angular speed $\omega$ in the presence of other static heavy quarks uniformly distributed over
$\mathcal{N} = 4$ SYM plasma. It has been pointed out in many occasions that in the course of rotational dynamics there is an
interference between the medium induced energy loss (drag) as well as the radiative energy loss
associated with the quark acceleration in the strongly coupled medium
\cite{Fadafan:2008bq, Chernicoff:2008sa, Hatta:2011gh, AliAkbari:2011ue, Kiritsis:2011ha}. The physical picture of
medium induced energy loss is related to the energetic collision and momentum transfer of the external probe with thermal plasma whereas
the radiative energy loss is nothing but the QCD realisation of Bremsstrahlung radiation. In this holographic study, we focus on the different range of angular speed ($\omega$) and the linear speed ($v = \mathcal{R} \omega $) of the probe quark to identify the regions of dominance of both drag and radiation.

It is interesting to note that unlike the heavy probe quark, the colour neutral $q\bar{q}$ bound states do not experience dissipative energy loss while performing a linear motion through the strongly coupled thermal plasma \cite{Peeters:2006iu, Chernicoff:2006hi,Chernicoff:2006yp, Athanasiou:2008pz}. In the dual gravity scenario, the motion of the probe string continues without being dragged and the string profile remains un-trailed \cite{Chernicoff:2006hi}.
In the present work, we holographically showed that for rotational dynamics of a heavy probe $q\bar{q}$ pair the dual string profile still remains unaffected from rotational drag.

The paper is organised as follows. In section \ref{jet}, we estimate the jet quenching parameter. We then compute the screening length in section \ref{screen}. Section \ref{rotq} is devoted to the detailed discussion on
the energy loss of a heavy probe quark rotating in the back reacted plasma. In section \ref{rotqmesoncal}, we showed that the heavy rotating $q\bar{q}$ probe in the presence of a static heavy quark distribution is free of rotational drag. Finally, we conclude with the significance of our main results in section \ref{con}.

\section{ Jet quenching parameter}\label{jet}
Following the holographic prescription given in \cite{Liu:2006ug}, in this section
we compute the jet quenching parameter ($\hat{q}$) and study the effect of the back reaction on it. Phenomenologically, the parameter is related to the
energy loss due to the suppression of heavy quark with high
transverse momentum in the presence of thermal medium.

In field theoretic point of view, the connection between the jet quenching parameter
and the expectation value of light-like Wilson loop in the adjoint
representation is established in the following way \cite{Liu:2006he},
\begin{equation}\label{jetq}
\langle W^A({\cal C})\rangle =  e^{-\frac{1}{4\sqrt 2} \hat q L^- L^2}.
\end{equation}
Here, the Wilson loop $\cal C$ is traced out by the separation length $L$ of a $q\overline q$ pair and a length $L^-$ along the light cone of the boundary gauge theory.
Since the gauge theory is strongly coupled, computation of the expectation value of light-like Wilson loop is extremely difficult due to lack
of systematic formulation. However, within the domain of gauge/gravity correspondence, we can calculate the expectation value using the following holographic prescription,
\begin{equation}\label{expect1}
\langle W^F({\cal C})\rangle = e^{iS({\cal C})}.
\end{equation}
Here $S({\cal C})$ is the Nambu-Goto action for the fundamental string with two of it's endpoints attached to the boundary and are dual to the boundary quark-antiquark
pair. The string action can be written as,
\begin{equation}{\label{wsaction}}
S= -\frac{1}{2\pi\alpha'}\int d\tau d\sigma \sqrt{-{\rm det}\, g_{\alpha\beta}},
\end{equation}
where $\alpha'$ is related with the string tension and $g_{\alpha\beta}$ is the induced world-sheet metric,
\begin{equation}\label{indmetric1}
g_{\alpha\beta} = {\partial x^\mu\over\partial \xi^\alpha}{ \partial x^\nu\over \partial \xi^\beta}G_{\mu\nu}.
\end{equation}
Here $G_{\mu\nu}$ is the background metric given in the equation (\ref{dualgrav}) and $\xi^{\alpha}$ are the world sheet coordinates where $\alpha = 0,1$.

It is important to note that the Wilson loop considered in (\ref{expect1}), is in the fundamental representation. However,
by using the group theoretical identity, $Tr_{Adj}=Tr_{Fund}^2$, it is
easy to translate the form of expectation value from fundamental representation to adjoint
representation as,
\begin{equation}\label{expect2}
\langle W^F({\cal C})\rangle^2= \langle W^A({\cal C})\rangle.
\end{equation}
A combination of equations (\ref{jetq}), (\ref{expect1}) and (\ref{expect2})
leads to a holographic working formula for jet quenching parameter in the dual gravity theory,
 \begin{equation}
\hat q = -\frac{8\sqrt 2i}{L^{-} L^2} (S-S_0),
\end{equation}
where $S_0$ is the self energy contribution due to the total mass of $q$ and $\overline q$.
In order to compute the action $S$ it is customary
to write down the background metric (\ref{dualgrav}) in the light-cone coordinates. In this coordinates the metric becomes,
\begin{equation}
ds^2= f[-(1+h)dx^+dx^- +\frac{1}{2}(1-h)\{ dx^{+2}+dx^{-2}\} + dy^2 +  dz^2 + {du^2\over h}].
\end{equation}
In the above metric we assume the  definition of $x^{\pm}$ as follows,
\begin{equation}
x^{\pm} =\frac{t \pm x}{\sqrt 2}.
\end{equation}
We choose the static gauge, $\xi^0=x^-(L^-\geq x^-\geq 0),
\xi^1 =y(-\frac{L}{2}\leq y\leq \frac{L}{2} )$ and also set the $q\overline q$ pair at $y=\pm \frac{L}{2}$ on
$x^+ = {\rm constant}$, $z = {\rm constant}$ plane. With these choices of parameters and by considering
$L_{-} \gg L$, the profile of the string is entirely constrained to $u = u(y)$.
Consequently the string action of equation (\ref{wsaction}) takes the form as,
\begin{equation}\label{integrand}
S= \frac{iL^-}{\sqrt 2 \pi \alpha'}\int_0^{\frac{L}{2}}dy f\sqrt{(1-h)(1+\frac{u'^2}{h})}.
\end{equation}
The fact that the above form of action does not explicitly depend on $y$ leads to the following conservation equation,
\begin{equation}
\frac{\partial {\cal L}}{\partial u'} u' - {\cal L} = E.
\end{equation}
Here $E$ is the constant of motion and $\mathcal{L}$ is the integrand of equation (\ref{integrand}).
Finally the equation of motion for the $u$ can be written as,
\begin{equation}\label{uprime}
u'= \sqrt{h[\frac{f^2(1-h)}{E^2}-1]}.
\end{equation}
Now from the symmetry of the problem we identify the boundary conditions as,
$u(\pm \frac{L}{2}) = 0$ and $u'(0) = 0$.
The second boundary condition related to the extrema of the variable $u$ signifies the existence of physical turning point(s).  If we apply the boundary condition $u'(0) = 0$ in the equation (\ref{uprime}), we get two possible conditions for occurring turning point(s),
\begin{eqnarray}
 h = 0\quad {\rm and}\quad h =1-\frac{E^2}{f^2}.
\end{eqnarray}
Certainly, $h=0$ sets the turning point on the horizon $u_t^1 = u_+$ whereas the other condition $h =1-\frac{E^2}{f^2}$ , for small values of $E$, sets the other turning point $u_t^2 $ very close to the boundary. We also notice that the right hand side of equation (\ref{uprime}) remains positive very close to horizon and becomes negative near boundary.  The physical consistency demands ${u^{'}}^2 \geq 0$ always and that is not true in the range $ 0 \leq u< u_t^2$. So to avoid this region of inconsistency we set a cut-off on boundary at some radial value $ u = \delta > u_t^2$.
In this way we can safely set the non-negative ${u^{'}}^2$ in
the range $\delta < u < u_+$. In the end, we consider the limit $\delta \rightarrow 0$ to make the final result cut-off independent.
By using the form of $u^\prime$ of equation (\ref{uprime}), we can re-write the action for the fundamental string as,
\begin{equation}
S= \frac{iL^-}{\sqrt 2 \pi \alpha'}\int_\delta^{u_+} du f \sqrt{\frac{1-h}{h}} [1-\frac{u^4 E^2}{l^4(1-h)}]^{-\frac{1}{2}}.
\end{equation}
The above action is divergent since it contains self energies of the quark and antiquark pair.
The self energy contribution can be holographically realised by considering the world-sheets of two free
straight fundamental strings both hanging from the boundary
to the horizon. Within the choice of gauge $x^{-} = \xi^0, u = \xi^1$, the self-contribution reads as,
\begin{equation}
S_0 = \frac{iL^-}{ \pi \alpha'}\int_\delta^{u_+} du\sqrt{G_{uu} G_{--}}
=\frac{iL^-}{\sqrt 2 \pi \alpha'}\int_\delta^{u_+} du f\sqrt{\frac{1-h}{h}}.
\end{equation}
Now, the regularised action of our interest takes the following form,
\begin{equation}\label{sszero}
S-S_0 =\frac{iL^-}{\sqrt 2 \pi \alpha'}\int_\delta^{u_+} du f \sqrt{\frac{1-h}{h}}
\Big[\{1-\frac{u^4 E^2}{l^4(1-h)}\}^{-\frac{1}{2}}-1\Big]
\approx \frac{iL^-E^2}{2\sqrt 2 \pi \alpha'}I_1,
\end{equation}
where,
\begin{equation}
I_1= \int_\delta^{u_+}\frac{ du}{f\sqrt { h(1-h)}}.
\end{equation}
At this point we replace $E$ in terms of quark-antiquark pair separation distance $L$. In order to do so, we first compute the separation distance $L$ between the quark-antiquark pair from equation (\ref{uprime}) and it comes out as,
\begin{eqnarray}
 L = 2\int^{u_+}_{\delta} du \frac{1}{\sqrt{h\Big[\frac{f^2(1-h)}{E^2} -1 \Big]}}.
 \label{length}
\end{eqnarray}
For a given small separation between $q\bar{q}$ pair we can invert the above equation and estimate the conserved
parameter $E$ up to the first order in $L$ as,
\begin{equation}
E= \frac{L}{2I_1} - \frac{E^3}{2}\frac{I_2}{I_1} \approx \frac{L}{2I_1} + {\cal O}{(L)}^3,
\end{equation}
where we define,
\begin{equation}
I_2= \int_\delta^{u_+}\frac{ du}{f^3 (1-h)\sqrt {h(1-h)}}.
\end{equation}
Therefore we obtain,
\begin{equation}
S-S_0 \approx \frac{iL^-L^2}{8\sqrt 2 \pi \alpha' I_1},
\end{equation}
and correspondingly the jet quenching parameter takes the form as,
\begin{equation}
\hat q = \frac{1}{ \pi \alpha' I_1}.
\end{equation}
However, by using the relation $\frac{l}{\alpha'^2} = g_{YM}^2 N_c$ we can settle the form of the jet quenching parameter
in terms of boundary parameters,
\begin{eqnarray}
 \hat q = \frac{\sqrt{ g_{YM}^2 N_c}}{ \pi  I_1' (T,b)},
\end{eqnarray}
\begin{figure}[ht]
\centering
\mbox{\subfigure[Plot 1]{\includegraphics[width=6.5 cm]{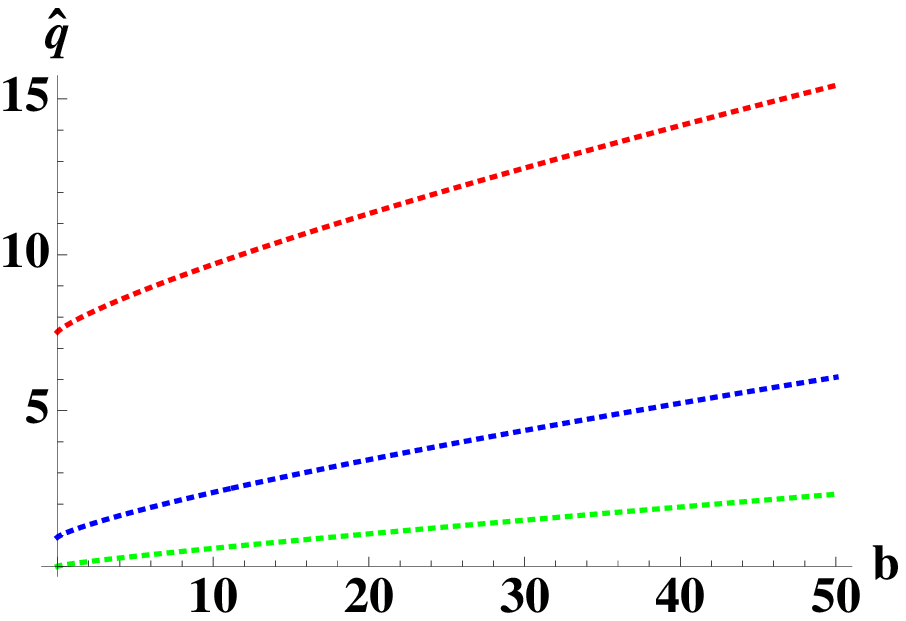}}
\quad
\subfigure[Plot 2]{\includegraphics[width=6.5 cm]{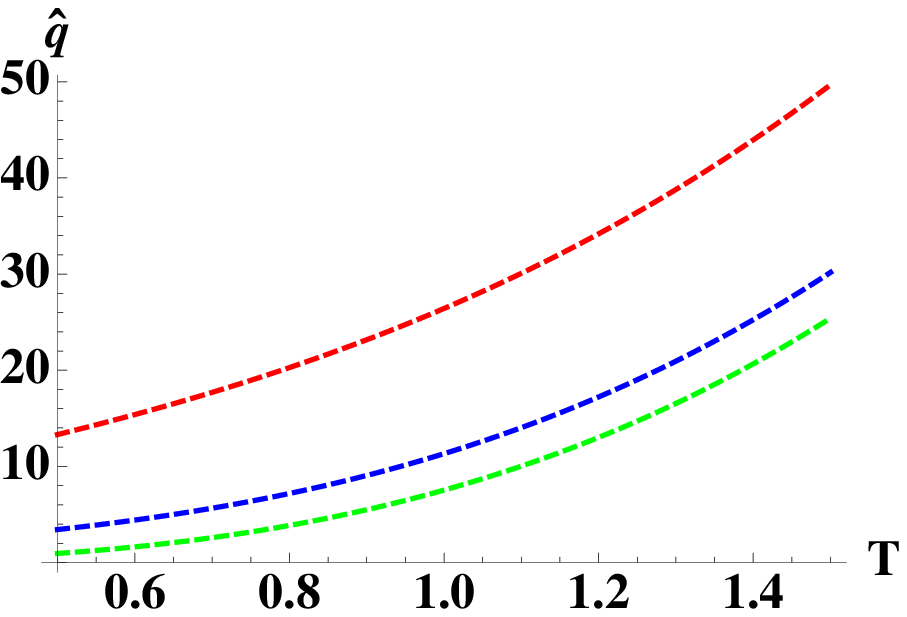} }}
\caption{Plot 1 shows the variation of jet quenching parameter as a function of quark density $b$ for
$T = $ 0.1(Green), 0.5 (Blue), 1 (Red) respectively.
Plot 2 describes the variation of
jet quenching parameter as a function of $T$ for the
values of $b= $ 0 (Green), 20 (Blue), 120 (Red) respectively.}
\label{figjet}
\end{figure}
where $I_1' (T,b) = \int_\delta^{u_+(T,b)}\frac{ u^2 du}{\sqrt { h(1-h)}}$ and $g_{YM}$ is the Yang Mills (YM) gauge coupling. To have a better understanding of the back reaction
effect on the jet quenching phenomenon we plot the relevant parameter $\hat{q}$ with respect to quark cloud density $b$, keeping temperature $T$ fixed (Plot 1).  We find that the parameter $\hat{q}$ increases monotonically as we tune up the value of quark density from zero to some finite number. This implies that the presence of heavy static quarks back-reacting the plasma enhances the energy loss due to the suppression of the external heavy probes moving with high transverse momentum. We also plot the parameter $\hat{q}$ with respect to temperature, $T$ keeping the quark cloud density fixed. We observe again that $\hat{q}$ monotonically increases with temperature. It is important to note that
at zero temperature the jet quenching parameter is finite and increases with respect to the magnitude of back reaction. To summaries we notice that the \textit{jet quenching phenomenon enhances as we increase the back reaction as well as the temperature of the plasma}.

\section{Screening Length}\label{screen}
The purpose of the present section is to study
the screening length ($L_s$) of a $q\bar{q}$ pair probing the back-reacted $\mathcal{N} = 4$
SYM plasma. Screening length ($L_s$) is defined as the maximum separation between a
$q\bar{q}$ pair moving with a constant speed in the plasma. If the separation
between them exceeds $L_s$, they get detached from each other with no binding energy. Consequently
they become screened in the QGP medium. The holographic computation of the screening length
is prescribed in \cite{Liu:2006nn} and the prescription requires the consideration of a  time-like Wilson
loop($\mathcal{C}_{time-like}$) traced out by the $q\bar{q}$ pair.
Moreover, the computation becomes much simpler in the rest frame of $q\bar{q}$ pair where plasma flows with a
constant speed. Correspondingly, in the dual theory, the black hole background is boosted by a
rapidity parameter. For the shake of holographic computation, in this boosted background, we consider a fundamental string with both of it's ends attached to the boundary of the space time. The end points of the fundamental string are realised as the holographic dual to the  $q\bar{q}$ pair in the boundary theory.  As the separation of $q$ and $\bar{q}$ approaches to the $L_s$, in the dual picture, the body of the string tends to reach at the horizon of the geometry.
When the separation goes beyond $L_s$, two isolated strings are energetically favourable in the dual theory.
Binding energy of
$q$ and $\bar{q}$
 is related to the thermal expectation
value of the time like Wilson loop operator, $\langle \mathcal{W}(\mathcal{C}_{time-like})\rangle$.
Thereby, using the holographic
mapping between $\langle \mathcal{W}^{fund}(\mathcal{C}_{time-like})\rangle$ and $e^{iS}$,
we calculate the binding energy in dual gravity.

The set up for holographic computation is followed by some assumptions in the dual boundary theory.
Firstly, we consider that in the rest frame of $q\bar{q}$ pair,
the thermal plasma moves along a flat boundary coordinate ($z$) with a constant speed $v$ .
We also assume that the Wilson loop traced out by the $q\bar{q}$ pair lies in the $t-x$ plane.
We specify the temporal length and the spatial length of the loop by the parameters
$\mathcal{T}$ and $L$ respectively. Finally we consider the limit $\mathcal{T} \gg L$
signifying the invariance of string world-sheet under time translation.

For the shake of holographic computation of $L_s$, we introduce a boost in the dual gravity background in the following way,
\begin{eqnarray}
dt&=& \cosh\eta \, dt^* -\sinh\eta \, dz^*\nonumber\\
dz &=&  -\sinh\eta \, dt^* + \cosh\eta \, dz^*.
\end{eqnarray}
Under this boost the metric (\ref{dualgrav}) takes the following form,
\begin{eqnarray}
ds^2 &=&  f\big[-\{1-\cosh^2\eta (1-h)\}dt^{*2} + \{1+(1-h)\sinh^2\eta\}dz^{*2} \nonumber\\
&&-2 (1-h)\cosh\eta\sinh\eta\, dt^*dz^* + dx^{2} + dy^2 + \frac{du^2}{h}\big],
\label{boost}
\end{eqnarray}
where $\eta = \tanh^{-1}v$ is the rapidity parameter.
In the due course of computation we assume a choice of static gauge,
\begin{eqnarray}
 \tau = t^*, ~ \sigma=x, ~ y=z^*=0,
\end{eqnarray}
and a set of suitable boundary conditions,
\begin{eqnarray}
  u(\sigma = \pm \frac{L}{2})=0, ~ u(\sigma=0)= u_{extrm}, ~ u'(\sigma=0) = 0.
\end{eqnarray}
With the above choice of static gauge and boundary conditions the world-sheet action of equation (\ref{wsaction}) for dual fundamental string reads as,
\begin{equation}\label{screenaction}
S= -\frac{\mathcal{T} ~ l^2}{2\pi \alpha'}  \int d \sigma\sqrt{\frac{1}{u^4}\{1 - \cosh^2\eta(C_1 u^4 +
 C_2  u^3 )\}\{1 + \frac{ u'^2}{1 -C_1 u^4 -
 C_2  u^3 }\}},
\end{equation}
where $C_1$ and $C_2$ are defined as,
\begin{eqnarray}
 C_1 = \frac{2m}{l^6}, ~  C_2 = \frac{2b}{3l^4}.
\end{eqnarray}
The Lagrangian in the above string action does not explicitly depend on $\sigma$.
Consequently we can construct a Hamiltonian like function as a constant of motion,
\begin{eqnarray}\label{screeneqm}
 \frac{\partial \cal L}{\partial  u'} u' -{\cal L} = W.
\end{eqnarray}
It is straightforward to derive the equation of motion of $ u$ coordinate by combining equations (\ref{screenaction}) and (\ref{screeneqm}),
\begin{eqnarray}\label{screeneqm1}
  u'^2 = \frac{\{1 -C_1 u^4 -
 C_2  u^3 \}}{W^2  u^4}\{1 - \cosh^2\eta(C_1 u^4 +
 C_2  u^3 ) -W^2  u^4\}.
 \label{screenl}
\end{eqnarray}
Clearly, $u$ has an extrema ($ u_{extrm1}$) that lies on the horizon itself,
\begin{equation}
 1 -C_1 u_{extrm1}^4 -C_2  u_{extrm1}^3 =0.
 \label{extrm1}
\end{equation}
However, if the condition of extrema (\ref{extrm1}) is attained,
the string reaches at the horizon, breaks down into two separate strings, holographically corresponds to a pair of free quark with no binding energy. The other extrema ($u_{extrm2} $) is fixed by
the following constraint,
\begin{eqnarray}\label{screenconstrain}
 \frac{1}{W^2 }
\Big[ \frac{1}{ u_{extrm2}^4}\{1 -
\cosh^2\eta(C_1 u_{extrm2}^4 + C_2  u_{extrm2}^3 )\} -W^2 \Big] =0.
\end{eqnarray}
Since hyperbolic cosine function is always positive and  $\ge 1$ , therefore the factor $\frac{1}{W^2 }
\Big[ \frac{1}{ u^4}\{1 - \cosh^2\eta(C_1 u^4 + C_2  u^3 )\} -W^2 \Big]$ takes negative value and $ u'$ becomes unphysical in the range $u \in [ u_{extrm2}, u_{extrm1}]$. On the other hand, for sufficient small $W$, the factor reduces, 
\begin{eqnarray}
 \frac{1}{W^2}
\Big[ \frac{1}{ u^4}\{1 - \cosh^2\eta(C_1 u^4 + C_2  u^3 )\} -W^2 \Big] \nonumber
\approx \Big[\frac{1}{W^2 u^4}-1\Big]
\end{eqnarray}
and is always positive and large near the boundary. It is then natural to conclude that $ u'^2$
switches sign at $ u_{extrm2} =  u_c$ and we identify $ u_c$ as a physical turning
point of the string configuration.
By solving the constraint equation (\ref{screenconstrain}) we explicitly determine $ u_c$.
Once this physical turning point is extracted, integrating (\ref{screenl}) and exploiting the boundary condition
$u(\sigma = \pm \frac{L}{2})=0$ we obtain the separation distance between a $q\overline q$ pair,
\begin{equation}
L = 2W\int_0^{ u_c}\frac{ u^2 d  u}{\sqrt{(1 -C_1 u^4 -
 C_2  u^3)\{1 - \cosh^2\eta(C_1 u^4 +
 C_2  u^3 ) -W^2  u^4\}} }.
 \label{screen1}
\end{equation}
To examine the effect of back reaction on the separation distance of a $q\bar{q}$ pair we plot the
$L$ with respect to the constant of motion $W$, keeping the rapidity parameter $\eta$ fixed. It is evident from both (plot 3) and (plot 4) that there is no separation distance, $L$ when the constant of motion $W$ takes zero value. For finite $W$, $L$ increases monotonically till it attains the maximum value corresponding to a certain $W$ and then it falls of. 
The maximum value of the separation length $L = L_{max}$ signifies that beyond this value of $L$ there is no
solution of equation (\ref{screenconstrain}). Physically it means that the $q\bar{q}$ pair dissociates with no binding energy
if they are separated beyond $L_{max}$ and this maximum value of $L$ is recognised as the screening length $L_s$ associated
with the $q\bar{q}$ pair.
\begin{figure}
\centering
\mbox{\subfigure[Plot 3]{\includegraphics[width=6.5 cm]{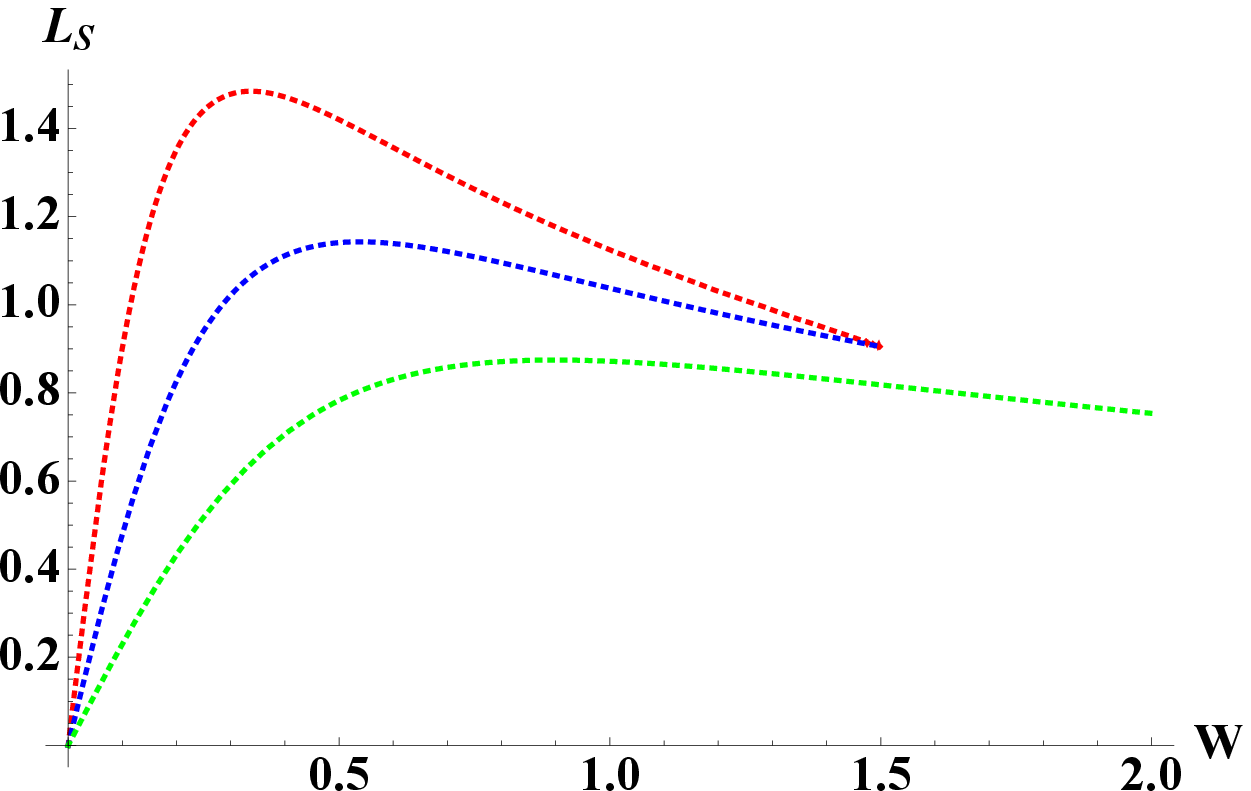}}
\quad
\subfigure[Plot 4]{\includegraphics[width=6.5 cm]{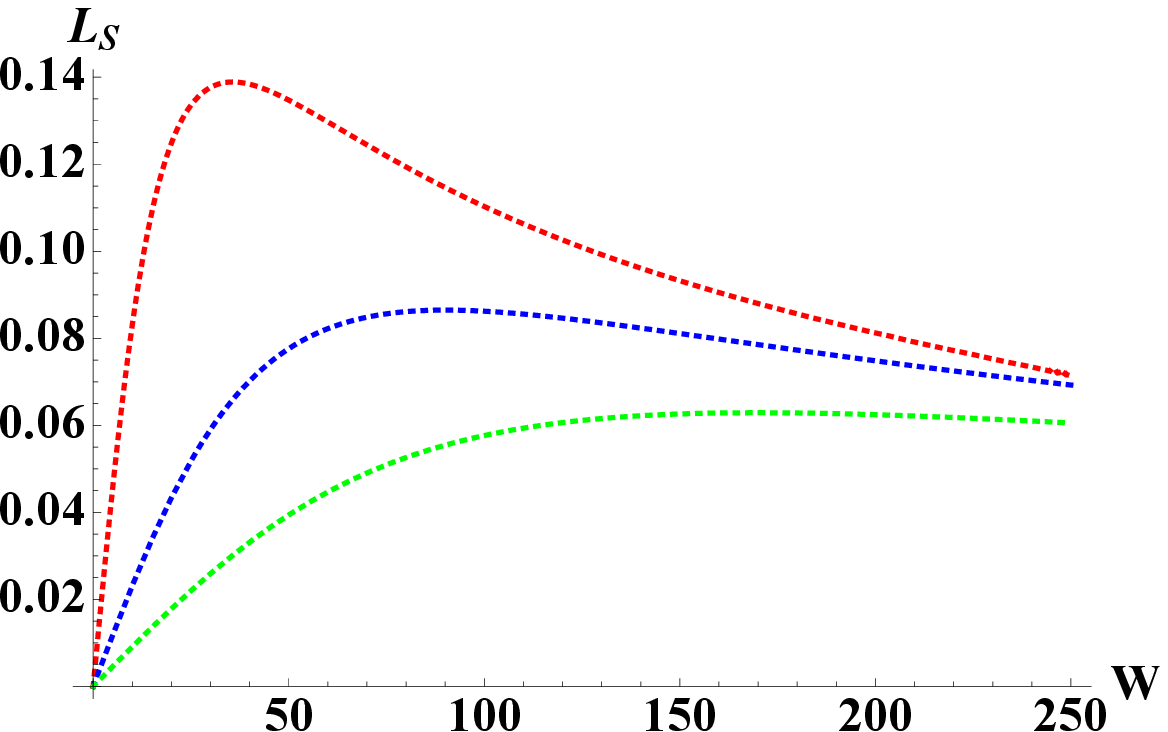} }}
\caption{Plot 3 shows the variation of screening length as a function of constant of motion for
$T = $ 1, $m=1$, $\eta = 1$ and $b = $1 (Red), $5$ (Blue), $15$ (Green) respectively.
Plot 4 describes the variation of screening length as a function of constant of motion for
$T = $ 1, $m=1$, $\eta = 5$ and $b = $1 (Red), $5$ (Blue), $15$ (Green) respectively.}
\label{figjet1}
\end{figure}
In (plot 3), we study the function
$L(W)$ for three different values of quark cloud densities (b=0, Red; b=1, Blue; b=10, Green)
and a fixed rapidity parameter $\eta = 1$. We find that \textit{the more
the plasma is back reacted, the less $L_s$ is allowed for a  $q\bar{q}$ pair}. In (plot 4) we consider the
same set values of quark density but fix the rapidity parameter at a higher value $\eta = 5$. Again we observe that
the enhancement of back reaction screens the $q\bar{q}$ pair at a lower value of the separation length .
However, for a fixed magnitude of back reaction, we find that $L_s|_{\eta =1} > L_s|_{\eta =5}$.

Furthermore, for a given set of values of $L < L_s$ there are two possible values of constant of motion $W$. Therefore to know the preferable one we find the minimum potential energy for a given set of values of $L < L_s$. 
The holographic computation of the binding energy $V$ is based on the consideration of the following
prescription,
\begin{equation}
V= - \frac{S-S_0}{\cal T},
\label{pot}
\end{equation}
where $S_0$ is the self energy contribution coming from two free quarks.  $S_0$ is realised as the Nambu-Goto action of a fundamental string hanging from the boundary to the horizon of the back reacted black hole geometry.  To compute $S_0$ we choose the gage as follows,
\begin{eqnarray}
\tau = t^{*}, ~ \sigma= u, ~ x = x(\sigma),~ y=0, ~ z^{*}=0
\end{eqnarray}
With this gauge choice, we compute the $S_0$ using following form of action,
\begin{eqnarray}
S_0 = -\frac{2 \mathcal{T}}{2 \pi \alpha'}\int_0^{u_{+}} \sqrt{-\text{det} g},
\end{eqnarray}
where $g_{\alpha\beta} = \partial_{\alpha}X^{\mu}\partial_{\beta}X^{\nu}G_{\mu\nu}$ is the induced metric and $G_{\mu \nu}$ is given in (\ref{boost}). The factor 2 comes in front of the action to take care of the contributions from both quark and anti-quark.


To study the potential energy between $q\bar{q}$ pair we use the equation (\ref{screen1}) to solve $W$ as a function of $L$
and plug back the solution $W(L)$ in (\ref{pot}). We then study the potential energy $V(L)$ of a $q\bar{q}$ pair as function of
separation distance between them.

\begin{figure}
\centering
\mbox{\subfigure[Plot 5]{\includegraphics[width= 6.5 cm]{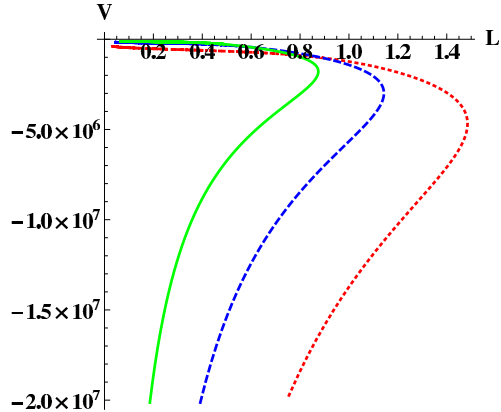}}}
\caption{Plot 5 shows the variation of binding energy $V$ of $q \bar{q}$ pair as a function of separation length $L$ between them. Here we study
the variation of potential energy for following three choices of parameters: red ($\eta = 5$, $T=1$, $b=1$); blue ($\eta = 5$, $T=1$, $b=5$) and green ($\eta = 5$, $T=1$, $b=15$).}
\label{figjet2}
\end{figure}
We plot the binding energy $V$ of a  $q\bar{q}$ pair as a function of $L$ for a fixed value of the rapidity parameter, $\eta = 5$. It is evident in this plot that for a given set of values of $L < L_s$ there are two possible branches of corresponding
binding energies. One branch is associated with the higher energy values (higher $W$) whereas the other branch corresponds
to the lower energy values (lower $W$) and it is the physically favourable energy configuration of
a $q\bar{q}$ pair. If the energy configuration of a $q\bar{q}$ pair is in the high energy branch, due to instability, the pair
makes a transition to the low energy branch. From the physically favoured lower branch of energy configuration, it is clearly visible that the \textit{the presence of the back reaction actually reduces the binding energy of a $q\bar{q}$ pair.}

\section{Energy loss of a rotating heavy quark}\label{rotq}
In this section we study the dynamics of a heavy probe quark rotating with a constant angular speed in the presence of
other static heavy quarks uniformly distributed over $\mathcal{N} = 4$ SYM plasma. In particular, we assume that the quark rotates on a two dimensional flat space along
a circle of radius $\mathcal{R}$ with a constant angular frequency $\omega$. The constant speed and
the constant acceleration related to the rotating quark
are given as $v=\mathcal{R} \omega$ and $a= \omega^2 \mathcal{R}$ respectively.
The strong interaction between the rotating quark and the back reacted thermal plasma results into an energy loss
either in the form of radiation or due to in-medium dissipation. The energy loss due to the interaction between the probe quark and the strongly coupled thermal plasma is very difficult to compute in the boundary theory. However, following the holographic methods described in \cite{Athanasiou:2010pv, Herzog:2007kh}, we can compute the same energy loss by studying the motion of a rotating spiral string probing the deformed AdS black hole space-time (\ref{dualgrav}) in the weakly coupled dual gravity theory. One of the end points of this spiral
string is attached to the boundary of this back ground geometry and holographically corresponds to the boundary rotating quark.
The body of the string experiences a
centrifugal force, takes a spiral profile and stretches up to the black hole horizon $u=u_+$. To achieve a holographic estimation of the energy loss in the bulk theory we need to study the dynamics of the rotating string governed by the Nambu Goto action (\ref{wsaction}).
Since we have assumed the quark in the boundary theory is constrained to rotate on a plane, in the dual gravity theory, we choose the following parameterizations of the
string world sheet preserving the $SO(2)$ symmetry,
\begin{equation}\label{rgauge}
X^\mu(\tau,\sigma)=(t=\tau, u=\sigma, x= \rho(\sigma)\cos(\omega t+\theta(\sigma)),
y=\rho(\sigma)\sin(\omega t + \theta(\sigma)), z=0).
\end{equation}
The parameters $\rho(\sigma)$ and $\theta(\sigma)$ are introduced to depict the radial and angular profiles of the rotating string and they obey
the following boundary conditions,
\begin{eqnarray}
 \rho(0) = \mathcal{R}, ~~~ \theta(0) = 0.
\end{eqnarray}
If we use the suitable ansatz (\ref{rgauge}) in the Nambu Goto action (\ref{wsaction}),
the Lagrangian density takes the following form,
\begin{equation}\label{rlag}
{\cal L}= [\frac{f^2}{h}(h-\rho^2\omega^2)+f^2(h-\rho^2\omega^2)\rho'^2+hf^2\rho^2\theta'^2]^{\frac{1}{2}},
\end{equation}
where, the prime denotes the derivative with respect to $u$. It is important to note that the Lagrangian density does not explicitly depend on the $\theta$ coordinate, so the conjugate momentum will be a constant of motion and can be written as,
\begin{equation}\label{eqnmtheta}
\Pi_\theta = \frac{\partial {\cal L}}{\partial\theta'}= \frac{hf^2\rho^2\theta'}{\cal L}.
\end{equation}
Rewriting (\ref{eqnmtheta}) according to our convenience we get,
\begin{equation}
\theta' = \sqrt{\frac{(h-\rho^2\omega^2)(1+h\rho'^2)}{h^2\rho^2(hf^2\rho^2-\Pi_\theta^2)}}
\label{eqnmtheta1}.
\end{equation}
Again it is important to note that the numerator under square root contains the factor $(h-\rho^2\omega^2)$
which is always positive at the boundary since the speed of the quark is always less than the speed of light. On the other hand, the same factor is negative at the horizon as $h(u_+) = 0$.
Therefore in between the boundary and the horizon, numerator changes sign at some special value of the radial coordinate. We consider this special value as the critical point ($u_c$) in the bulk. In order to avoid the imaginary value of $\xi'$, we insert the following conditions at this critical point,
\begin{eqnarray}
h(u_c)-\rho(u_c)^2\omega^2=0,\\
h(u_c)f(u_c)^2\rho(u_c)^2-\Pi_\theta^2 = 0.
\label{critrot}
\end{eqnarray}
By solving the above two equations we get,
\begin{eqnarray}
\rho(u_c)= \sqrt{\frac{\Pi_\theta}{f(u_c)\omega}},\\
f(u_c)= \frac{\Pi_\theta \omega}{h(u_c)}.
\label{critconstraint}
\end{eqnarray}
So the spiral profile of the rotating string starts at $u=0$, $\rho=\mathcal{R}$,
goes through $(u_c,\rho_c)$ and tends to be extended up to the black hole horizon.
It has been shown in \cite{Fadafan:2008bq} that the body of the string embedded in the range
$u <u_c$ are causally disconnected from the part of string embedded in $u>u_c$.
The physically relevant part of the rotating string is confined to the region $u <u_c$ and moves with a speed slower than the
local speed of light. It is important to note that $h(u_c)=\rho(u_c)^2\omega^2$ is the curve that signifies the radial profile $\rho_{l}$ of a
string moving with a local speed as same as that of light. All curves corresponding to the general radial profile of string moving
with speed slower than speed of light should intersect the curve $\rho_l$ at the critical point $u=u_c$.
To obtain the spiral profile of the rotating string we follow the strategy prescribed in \cite{Fadafan:2008bq}.
First, by using (\ref{eqnmtheta1})
we eliminate the $\theta'$ from the
equation of motion for $\rho$ coordinate. Consequently, the equation of motion for $\rho$ takes the form as,
\begin{eqnarray}
&&2(\Pi_\theta^2 - \omega^2 f^2 \rho^4)
-[2 f h \rho^3 (h - \omega^2 \rho^2)f' - \rho \{\Pi_\theta^2 - f^2 (2 h \rho^2 - \omega^2 \rho^4)\}h']\rho'\nonumber\\
&&+2 h (\Pi_\theta^2 - \omega^2 f^2 \rho^4)\rho'^2
-\rho^3 [2 f h^2 (h - \omega^2 \rho^2)f' - (\Pi_\theta^2 \omega^2 - f^2 h^2)h']\rho'^3 \nonumber\\
&&+2 \rho (h - \omega^2 \rho^2) (\Pi_\theta^2 - f^2 h \rho^2)\rho'' =0.
\label{eomrho}
\end{eqnarray}
Then we solve the differential equation (\ref{eomrho}) with appropriate boundary conditions
for $\rho$ as a function of $u$ and constant $\Pi_{\theta}$. Moreover by substituting the solution in
(\ref{eqnmtheta1}) and then integrating, the angular profile $\theta(u)$ can be obtained. Here we are interested to extract the radial profile
only as we will see later the measure of radial profile at the boundary has a direct consequence to estimate the rate of energy loss of the
boundary quark. However, to achieve an analytic solution for the equation of motion
for $\rho$ is extremely difficult except few occasions \cite{Athanasiou:2010pv}.

Instead of achieving an analytic solution here we solve (\ref{eomrho}) using numerical methods. We specify the boundary conditions by
fixing the values of $\rho$ and $\rho'$ at $u = u_c$ for a suitable choices of $\omega$ and $\Pi$. To determine  $\rho'(u_c)$ we expand the
radial coordinate $\rho(u)$ around $u_c$ and consider the terms up to linear order.
\begin{eqnarray}
 \rho(u) = \rho(u_c)+ \rho'(u_c) (u-u_c)+ \cdot\cdot\cdot.
 \label{expansion}
\end{eqnarray}

By plugging the expansion (\ref{expansion}) into (\ref{eomrho}) and keeping terms up to linear order in $u$
we observe that the zeroth order coefficient turns out to be zero to satisfy the constraints (\ref{critconstraint}), whereas the linear order
coefficient sets a quartic equation in $\rho'(u_c)$. The physically consistent solution of $\rho'(u_c)$ signifying the fact that the radius of
rotation is always real, positive and smaller than the critical radius is given as,
\begin{eqnarray}
 \rho'_c &=& \frac{1}{(4 u_c h_c^3 \rho_c)} \Big[-4 u_c^2 h_c^2 -h_c^3 \rho_c^2 + \omega^2 h_c^2 \rho_c^4 +
 2 u_ch_c^2 \rho_c^2 h'_c - \Pi^2 u_c^4 {h'_c}^2 \\ \nonumber
 && + {\Big( 16 u_c^2 h_c^5 \rho_c^2 + {(h_c^3 \rho_c^2 + \Pi^2 u_c^4 {h'_c}^2  +
      h_c^2 (4 u_c^2 - \omega^2 \rho_c^4 - 2 u_c \rho_c^2 h'_c))}^2)}^{\frac{1}{2}}\Big],
\end{eqnarray}
where we have assumed $\rho(u_c) = \rho_c$ and $h_c= h(u_c)$. However, $u=u_c$ is a singular point
for both the Lagrangian (\ref{rlag}) and the equation of motion (\ref{eomrho}). Furthermore, at $u=u_c$ the numerical method breaks down.
To overcome this difficulty we separately solve the equation (\ref{eomrho}) in the ranges defined from $u_c-\delta$ to the boundary as well
as from $u_c+\delta$ to the horizon and then combine them in a consistent way by considering $\delta \rightarrow 0$ limit.
\begin{figure}
  \centering
  \mbox{\subfigure[Plot 6]{\includegraphics[width=4.5 cm]{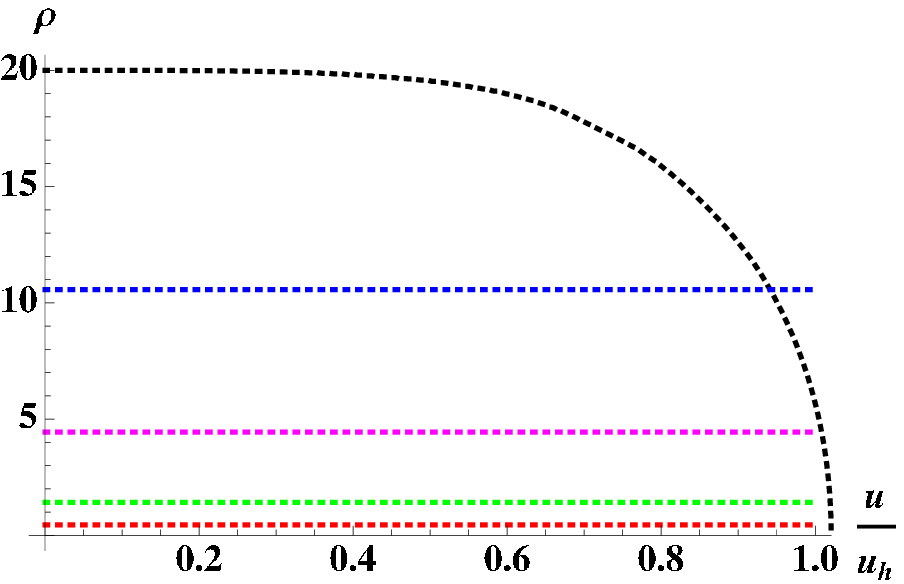}}
  \quad
  \subfigure[Plot 7]{\includegraphics[width=4.5 cm]{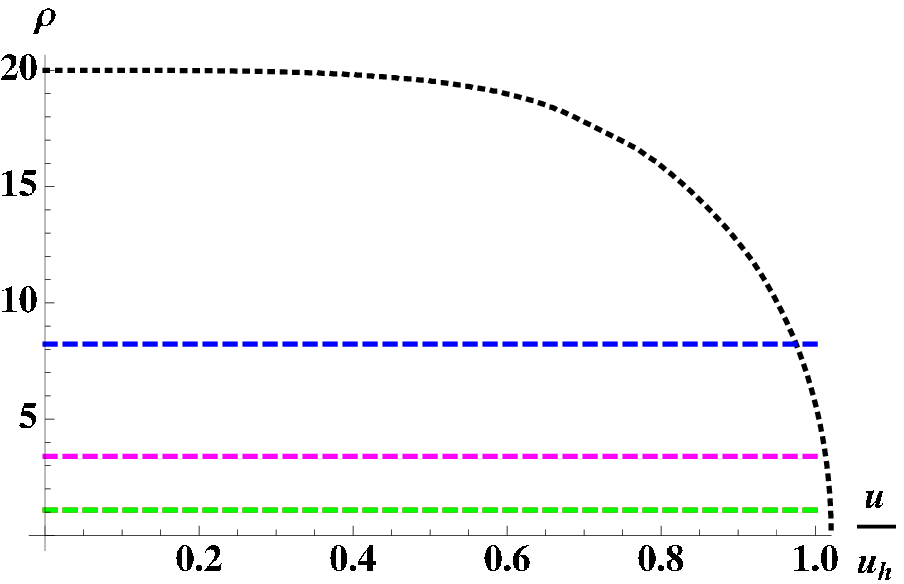} }
   \quad
  \subfigure[Plot 8]{\includegraphics[width=4.5 cm]{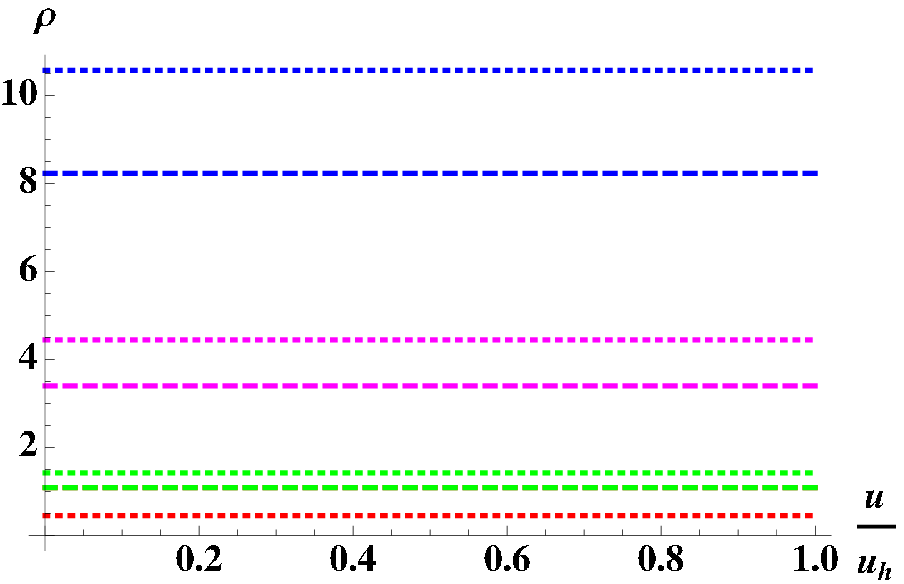} }}
 \caption{Plot 6, 7 show the radial dependence $\rho (u)$ of the rotating string for two different choices of densities of quark cloud,
 $b = .1 ( \text{plot} 6 ) $, $b=100 ( \text{plot} 7 )$ and for some fixed values of temperature $T = 1$ and $\omega  = .05 $.
 Each plot has four different branches corresponding to the four different values of
 momentum $\Pi_{\theta} = .1 $(red), $\Pi_{\theta} = 1 $(green), $\Pi_{\theta} = 10 $(pink) and $\Pi_{\theta} = 70 $(blue) respectively. In plot 8
 we compare the radial profiles for two different values of quark cloud density (dotted for $b = .1$ and dashed for $b = 100$).}
 \label{figrotq}
  \end{figure}

\begin{figure}
  \centering
  \mbox{\subfigure[Plot 9]{\includegraphics[width=4.5 cm]{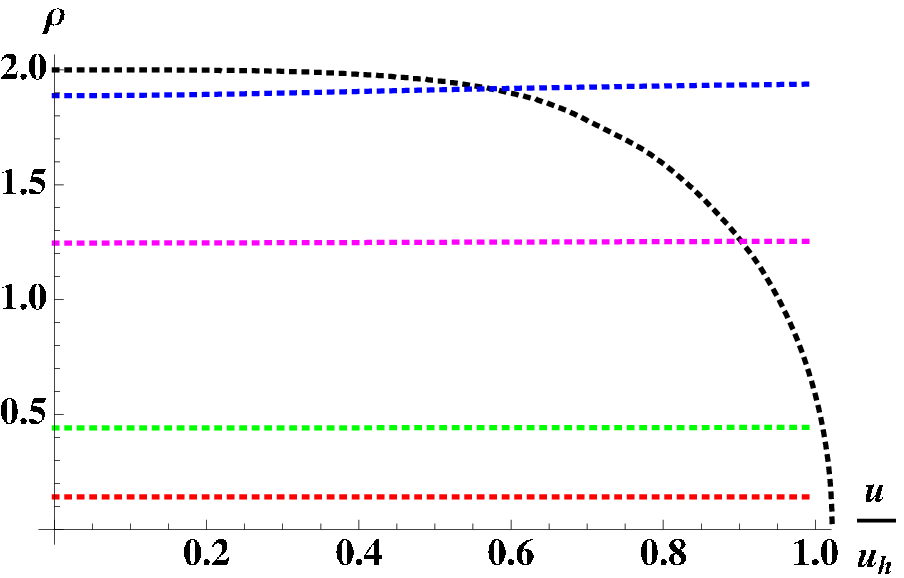}}
  \quad
  \subfigure[Plot 10]{\includegraphics[width=4.5 cm]{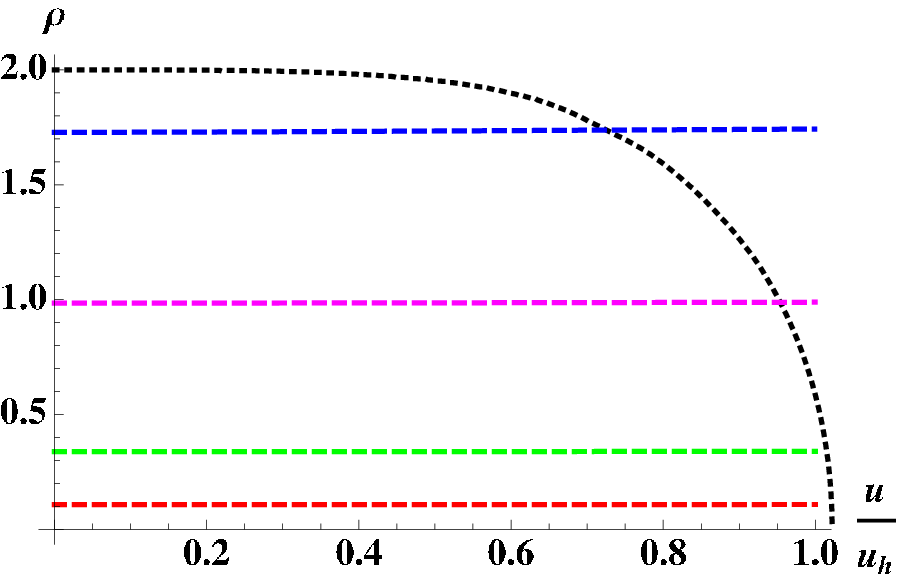} }
   \quad
  \subfigure[Plot 11]{\includegraphics[width=4.5 cm]{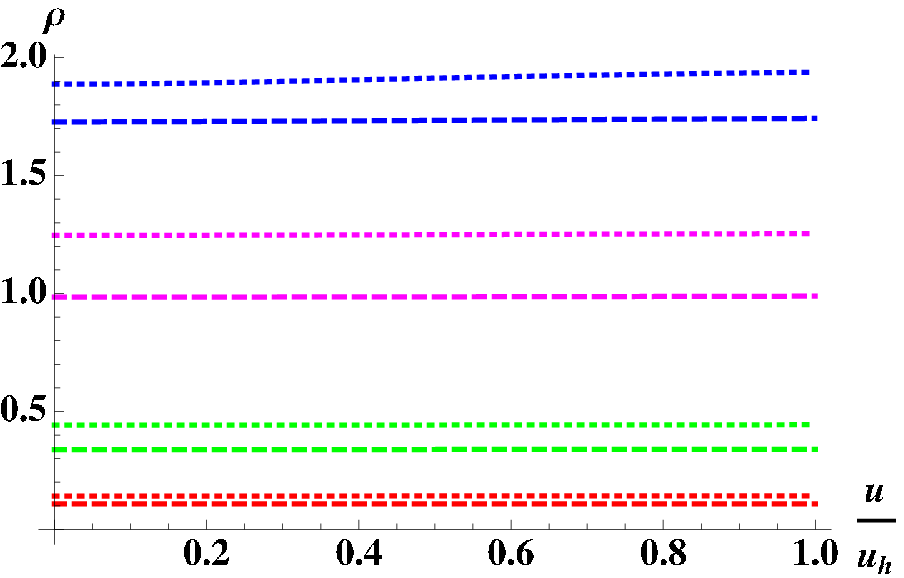} }}
 \caption{Plot 9, 10 show the radial dependence $\rho (u)$ of the rotating string for two different choices of densities of quark cloud,
 $b = .1 ( \text{plot 9}) $, $b=100 ( \text{plot 10})$ and for some fixed values of temperature $T = 1$ and $\omega  = .5 $.
 Each plot has four different branches corresponding to the four different values of
 momentum $\Pi_{\theta} = .1 $(red), $\Pi_{\theta} = 1 $(green), $\Pi_{\theta} = 10 $(pink) and $\Pi_{\theta} = 70 $(blue) respectively. Plot 11
 corresponds to a comparison between the radial profiles for two different values of quark cloud density (dotted for $b = .1$ and dashed for $b = 100$).}
 \label{figrotq11}
  \end{figure}

\begin{figure}
  \centering
  \mbox{\subfigure[Plot 12]{\includegraphics[width=4.5 cm]{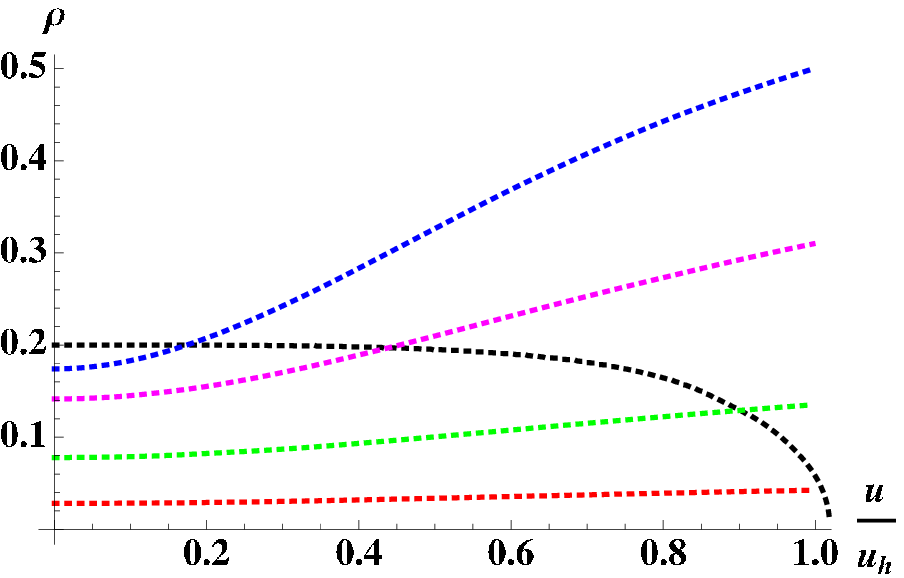}}
  \quad
  \subfigure[Plot 13]{\includegraphics[width=4.5 cm]{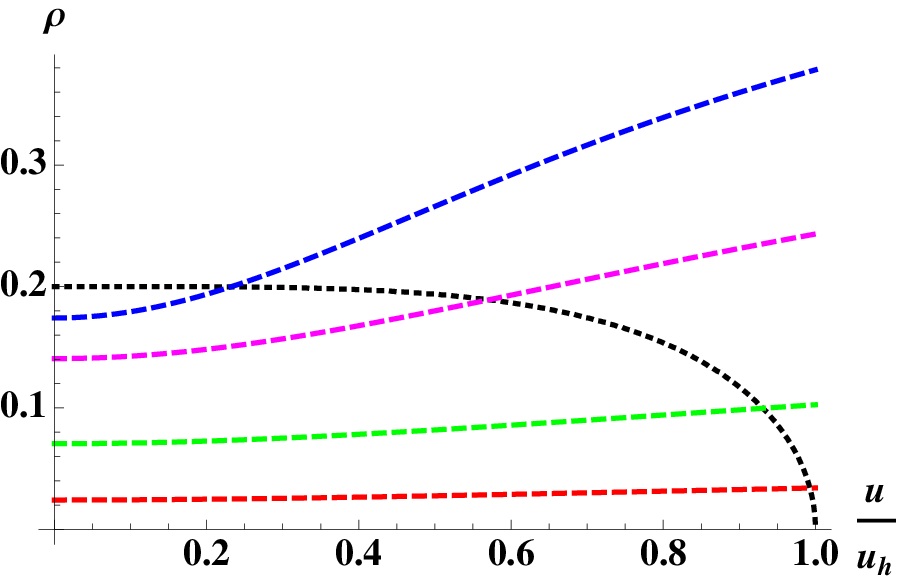} }
   \quad
  \subfigure[Plot 14]{\includegraphics[width=4.5 cm]{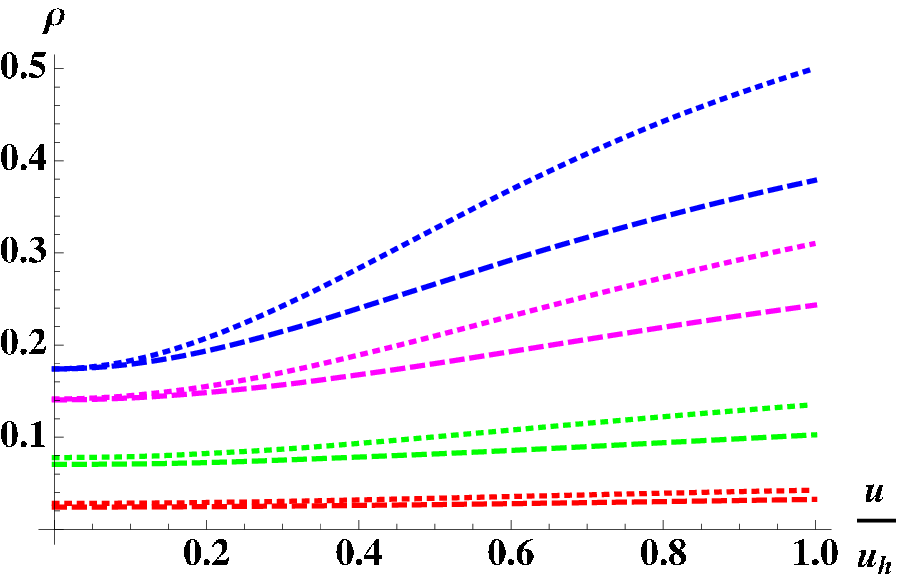} }}
 \caption{Plot 12, 13 show the radial dependence $\rho (u)$ of the rotating string for two different choices of densities of quark cloud,
 $b = .1 ( \text{plot 14}) $, $b=100 ( \text{plot 15})$ and for some fixed values of temperature $T = 1$ and $\omega  = 5 $.
 Each plot has four different branches corresponding to the four different values of
 momentum $\Pi_{\theta} = .1 $(red), $\Pi_{\theta} = 1 $(green), $\Pi_{\theta} = 10 $(pink) and $\Pi_{\theta} = 70 $(blue) respectively.  In plot 14
 we compare the radial profiles for two different values of quark cloud density (dotted for $b = .1$ and dashed for $b = 100$).}
 \label{figrotq12}
  \end{figure}

In figures (\ref {figrotq}), (\ref{figrotq11}), (\ref{figrotq12})  we notice that the radial profile of the rotating spiral
string is  characterised by  different choices of temperature $T$, string density $b$, conserved string momentum $\Pi_{\theta}$ and the angular speed $\omega$. For each profile there exists a unique limit $\rho(u \rightarrow 0) = \mathcal{R}$ holographically signifying the radius of the rotating quark in the boundary theory.

The fact that the speed of the boundary quark never exceeds the
speed of light put some constraint $\mathcal{R} < \omega$. Each radial profile clearly
validates the constraint $\rho(u \rightarrow 0) \omega < 1$ even if we increase the conserved string momentum $\Pi_{\theta}$ in an unbound way.  The intersection between the black dotted profile ($\rho_l$) and each of the radial profiles ($\rho(u)$)
fixes the value of the radius at the turning point $\rho(u_c)$. For a given $\omega$, as we increase the value of $\Pi_{\theta}$ the value
of $\rho(u_c)$ also gets enhanced. It is also evident from the plots that as $\omega < 1$, the radial profile is almost constant ($\rho(u=0) \approx \rho (u_c) \approx \rho(u=u_{+})$)
whereas for $\omega > 1$ the bending of string profile is significant and the radius at horizon is always bigger than the radius at boundary
($\mathcal{R}<\rho(u=u_+)$). For a given choice of momentum $\Pi_{\theta}$ and angular frequency $\omega$, if we increase the intensity of
back reaction the radius of rotation decreases accordingly. However this effect is more visible for $\omega < 1$.

Having discussed the generic features of the radial profile of the rotating string, now we study the a holographic
estimation of the rate of energy loss of a heavy probe quark rotating in the back reacted $\mathcal{N} =4$ SYM plasma.
 In the dual gravity theory, the holographic definition of the rate of energy loss associated
with the rotating string can be presented in the following form,
 \begin{equation}
{dE\over dt} =  - \frac{\delta S}{\delta(\partial_{\sigma}X^{0})} = \Pi^\sigma_t,
\end{equation}
 where $S$ stands for the Nambu Goto action.  Using the metric of the back reacted background (\ref{dualgrav}) in the above formula
 and the equation (\ref{critrot}) we re-write the expression for $\frac{dE}{dt}$ in the following way,
 \begin{eqnarray}
  {dE\over dt} = {hf^2 \omega \rho^2 \theta'\over 2\pi \alpha'\sqrt{-g}} = \frac{\Pi_{\theta} \omega}{2\pi \alpha'} =
  \frac{f(u_c)h(u_c)}{2\pi \alpha'}= \frac{h(u_c)}{2\pi \alpha' u_c^2}.
  \label{finen}
 \end{eqnarray}
 
 Here we consider $l=1$. Therefore the energy loss of a rotating string depends on the critical value $u_c (\Pi_{\theta}, \omega, b)$. However, to understand the influence
of the back reaction on the energy loss we prefer to study ratio between the energy loss with finite valued quark density and the same
with zero quark density with respect to the boundary quark speed $v$.
The holographic recipe to compute the aforementioned ratio is the following. First, we choose a set of $\omega$'s and $b$'s
and for each combination of $\omega, b $ we select a range of values for $\Pi_{\theta}$. For each values of $\Pi_{\theta}$ together with $\omega$ and $b$
we figure out the $(\rho_c, {\rho'}_c)$ and use them to solve the equation (\ref{eomrho}) by numerical method. Then we set the speed by
taking the boundary limit ($u \rightarrow 0$) of the solution ($v = \rho(u \rightarrow 0) \omega$).  For a fixed speed $v$, we holographically compute the rate of energy loss using equation (\ref{finen}). 

\begin{figure}[h]
  \centering
  \mbox{\subfigure[Plot 15]{\includegraphics[width=4.5 cm]{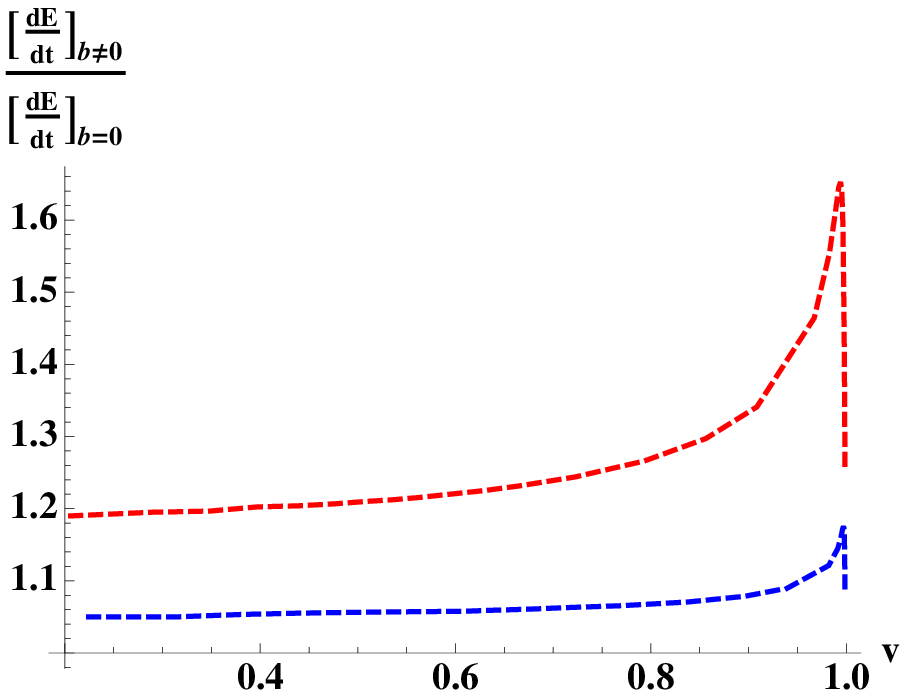}}
  \quad
  \subfigure[Plot 16]{\includegraphics[width=4.5 cm]{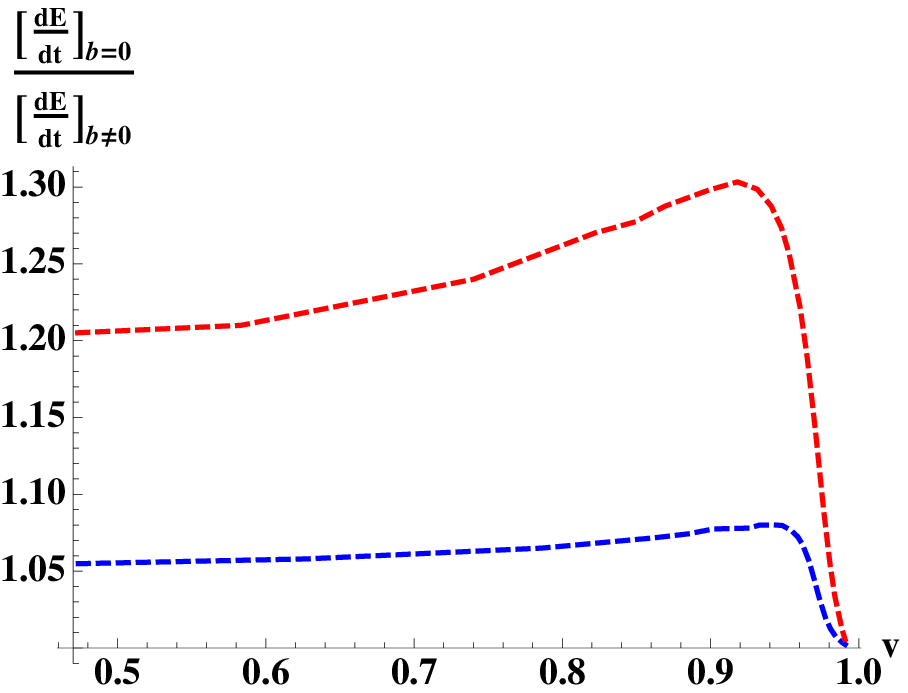} }
   \quad
  \subfigure[Plot 17]{\includegraphics[width=4.5 cm]{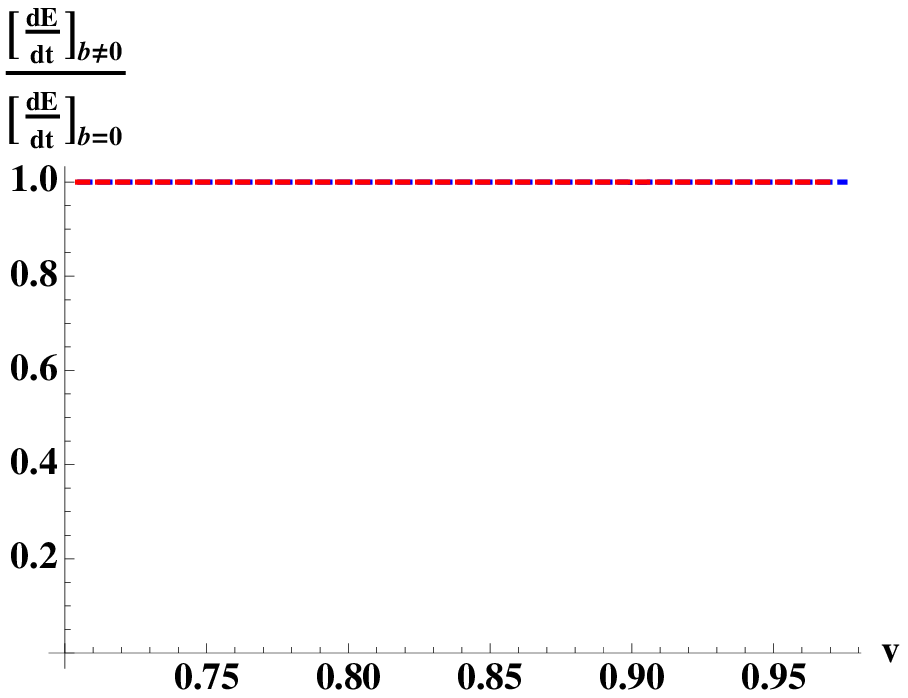} }}
 \caption{The ratio of the total energy loss in finite value quark density to the zero quark density for different angular velocities: $\omega=0.05, 0.5, 5.0$ from left to right. Each plot corresponds to two different values of quark density: $b=20$ (red), $b=5$ (blue) and temperature T=1.}
 \label{figrotq1}
  \end{figure}

In figure (\ref{figrotq1}) we plot the ratio between the rate of total energy loss in the back reacted SYM thermal plasma to the rate of total energy loss in the usual SYM thermal plasma as a function of speed. It is evident from the plot that \textit{the effect of back reaction enhances the energy loss due to the strong interaction between probe and the plasma}. For a lower and an intermediate angular speed, the  ratio $\frac{\frac{dE}{dt}|_{b\ne0}}{\frac{dE}{dt}|_{b=0}}$ increases monotonically and falls down to unity when the linear speed of probe approaches unity.  The fall of the ratio is more sharp for lesser value of the angular speed.  However, \textit{the back reaction effect ceases to exist for high values of angular speed.}

\begin{figure}[h]
  \centering
  \mbox{\subfigure[Plot 18]{\includegraphics[width=4.5 cm]{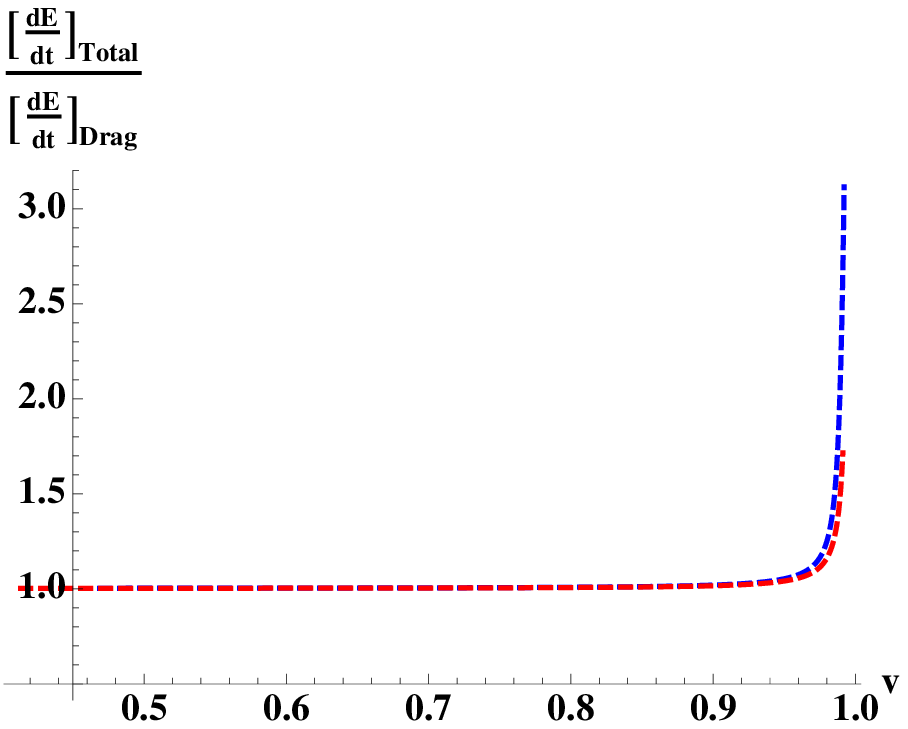}}
  \quad
  \subfigure[Plot 19]{\includegraphics[width=4.5 cm]{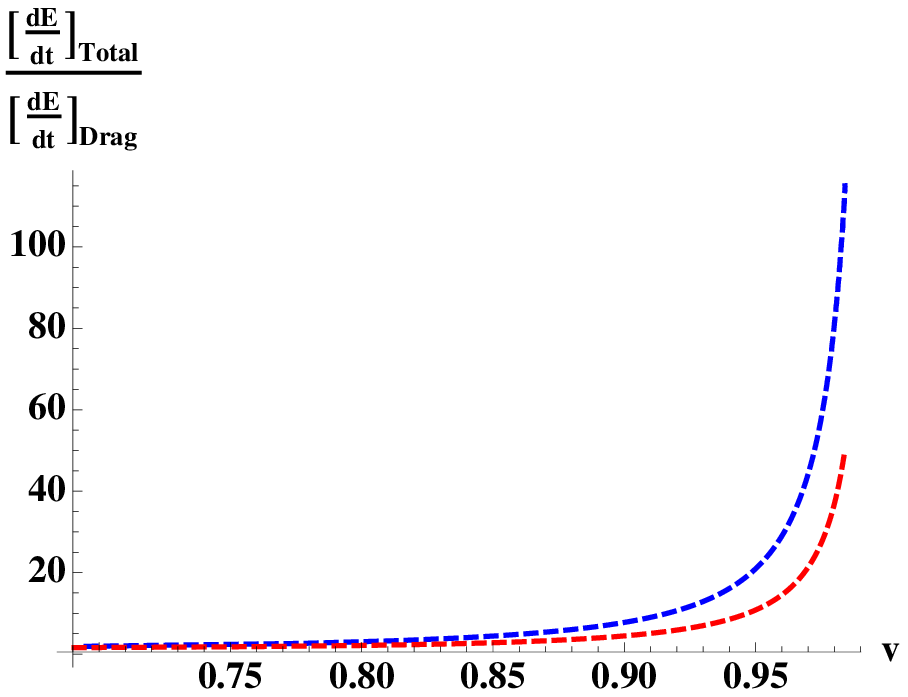} }}
 \caption{The ratio of the total energy loss rate to the drag energy loss rate for finite valued quark density and fixed temperature $T=1$.  The plots from left to right have different angular velocities; $\omega= 0.5$ and $5.0$. corresponds to two different values of quark density: $b=20$ (red), $b=5$ (blue). For $\omega = 5.0$, the numerical value of the ratio at the origin is unity.}
 \label{figrotq2}
  \end{figure}

In figure (\ref{figrotq2}) we plot the ratio between the total energy loss rate to the drag energy loss. The energy loss due to drag is given as,
 \begin{equation}
{dE\over dt}\Big|_{drag} =  - \frac{\delta S}{\delta(\partial_{\sigma}X^{0})} = \frac{h(u_c)}{2\pi \alpha' u_c^2}\Big|_{\text{drag}},
\end{equation}
where $u_c |_{\text{drag}}$ is the critical value of the radial coordinate when the string profile is trailed due to only drag. The profile of such string world sheet can be parameterized as follows,
\begin{equation}
X^\mu(\tau,\sigma)=(t=\tau, u=\sigma, x=vt + \xi(\sigma), y=0, z=0),
\end{equation}
where $\xi(u)$ is a function of radial coordinate signifying the trailing profile of the string.
With this gauge choice, the Nambu-Goto Lagrangian can be written as,
\begin{equation}
{\cal L}= -\frac{1}{2\pi\alpha'}\sqrt{f^2+hf^2\xi'^2-\frac{f^2v^2}{h}}.
\label{lag}
\end{equation}
Notice that the Lagrangian density (\ref{lag}) does not explicitly depend on $\xi$,
so the conjugate momentum, $\Pi_\xi $ for the field $\xi$ should be conserved and takes the form as,
\begin{eqnarray}
\Pi_\xi = -\frac{1}{2\pi\alpha'}{h f^2\xi'\over {\sqrt{f^2+hf^2\xi'^2-\frac{f^2v^2}{h}}}} = {\rm constant},
\label{mom1}
\end{eqnarray}
and some rearrangement of variables in the equation of motion with respect to the field $\xi$ gives,
\begin{eqnarray}
\xi' = \sqrt{\Pi_\xi^2(h-v^2)\over  h^2(\frac{hf^2}{4\pi^2 \alpha'^2}-\Pi_\xi^2)}.
\label{mom2}
\end{eqnarray}
The reality of $\xi'$ brings the imposition of the following constraints,
\begin{eqnarray}\label{constrain}
\nonumber h(u_c)&=&v^2,\\
\frac{h(u_c)f(u_c)^2}{4\pi^2 \alpha'^2} &=& \Pi_\xi^2.
\end{eqnarray}
where $u_c$ is the solution of the equation,
\begin{equation}
1-\frac{2mu_c^4}{l^6}-\frac{2}{3}\frac{bu_c^3}{l^4}-v^2=0.
\label{dragcrit}
\end{equation}

  \begin{figure}
  \centering
  \mbox{\subfigure[Plot 20]{\includegraphics[width=4.5 cm]{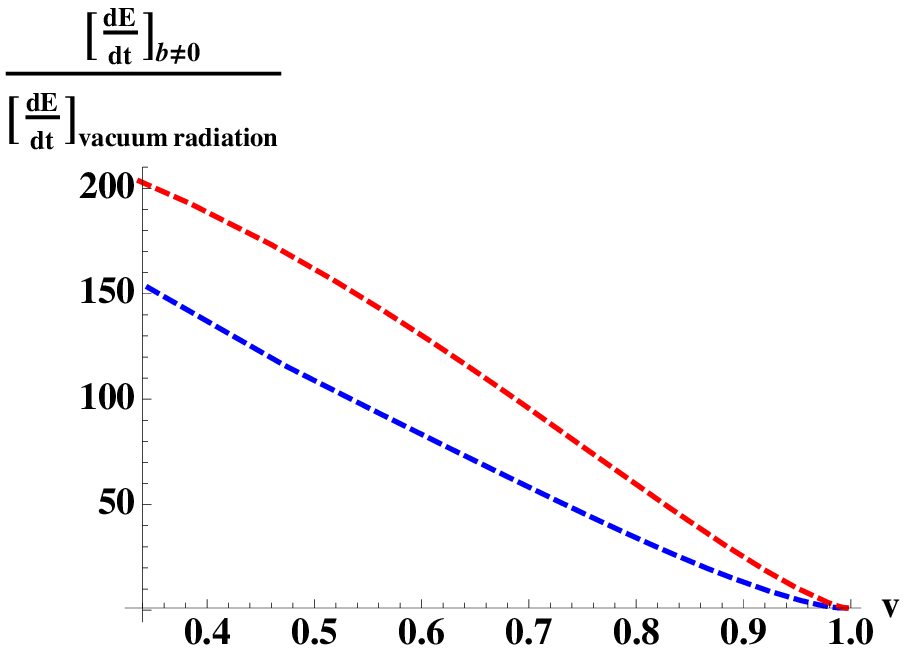}}
  \quad
  \subfigure[Plot 21]{\includegraphics[width=4.5 cm]{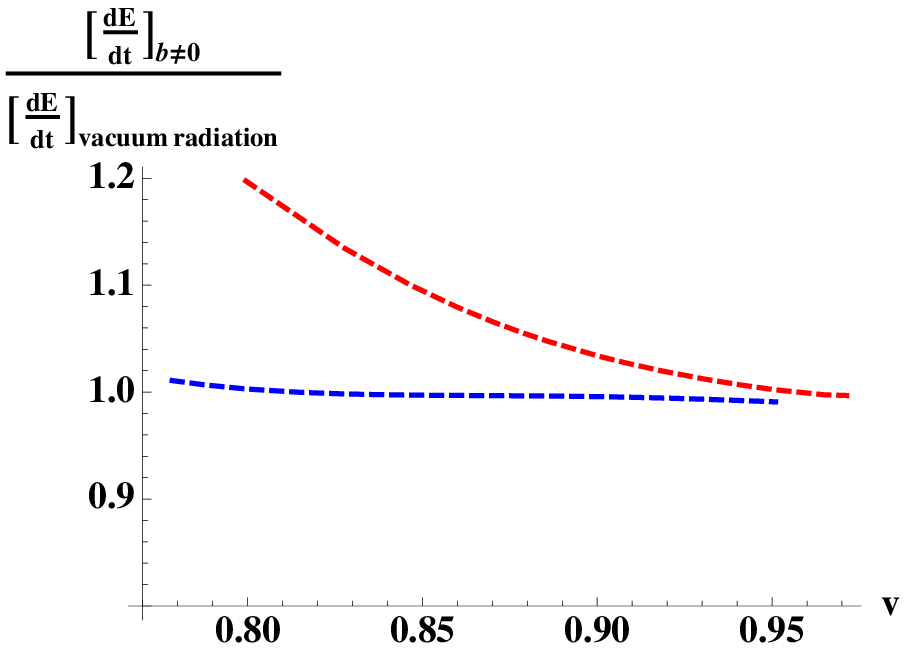} }}
 \caption{The ratio of the total energy loss rate in finite value quark density to the rate of vacuum radiation energy loss for fixed temperature T and different angular velocities: $\omega=0.5$, and $5.0$  from left to right. For $\omega = .05$, the numerical value of the ratio at the origin is unity. Each plot has two different values of quark density: $b=20$ (red)$b=5$ (blue) and a fixed temperature T=1.}
 \label{figrotq3}
  \end{figure}
It is evident from the plot (\ref{figrotq2}) that as angular speed is sufficiently small the total energy loss is dominated by the drag. The dominance of energy loss due to drag prevails even if the strength of back reaction takes lower value.  As the angular speed increases the ratio  $\frac{\frac{dE}{dT}|_{\text{total}}}{\frac{dE}{dT}|_{\text{drag}}}$ takes higher values than unity implying the fact that radiation energy loss contributes substantially. However, it is interesting to note that for {\it a higher value of angular speed, the more is the strength of back reaction the less is the contribution from radiation energy loss}.  It already evident from the plots (\ref {figrotq}), (\ref{figrotq11}), (\ref{figrotq12}) that at  small $\omega$ the speed of boundary quark is almost same as the local speed of the rotating string for each value of the radial coordinate. Therefore the corresponding string profile is very similar to dragged profile. However, for higher values of angular speed the influence of rotational motion modifies the string profile significantly.

Before closing this section we compare the total energy loss of a heavy probe quark performing rotational motion in a finite temperature back reacted plasma with the energy loss of a heavy probe quark rotating in the vacuum of the theory. The vacuum of the theory is realised as the  $\mathcal{N} = 4$ SYM theory. For pure rotational motion inside the strongly coupled thermal plasma, the vacuum energy loss of a boundary quark is first proposed by Mikhailov \cite{Mikhailov:2003er}. The form of the energy loss is given as,
\begin{equation}
\frac{dE}{dt}\Big |_{\text{vacuum radiation}} \sim \frac{v^2\omega^2}{(1-v^2)^2}.\label{mikh-formula}
\end{equation}
In (\ref{figrotq3}), we plot the ratio of total energy loss in the back reacted thermal plasma to the vacuum energy loss. We notice that for two different values of quark's angular speed 
the energy loss in back reacted thermal plasma never turns out to be lesser than the vacuum energy loss. The ratio smoothly falls off to unity as the quark's speed approaches to the speed of light. This property holds true even the strength of back reaction increases. By studying the plots, we infer that for a lower value of angular speed of the boundary probe quark, the dominating contribution for energy loss comes from the drag. However, for a higher value of the  angular speed the ratio is very close to unity. Therefore as the angular speed becomes substantially large the radiation energy loss starts dominating over drag energy loss. 

\section{Effect of angular drag on rotating heavy $q\bar{q}$ probe}\label{rotqmesoncal}

In this section we briefly study the effect of angular drag force on rotational motion of the heavy $q\bar{q}$ probe moving inside the back reacted thermal $\mathcal{N} = 4$ SYM plasma. In\cite{Chernicoff:2006hi}, it has been shown that the translational degrees of freedom of the $q\bar{q}$ probe are free of drag effect. In this present work, we show that the rotational degrees of freedom of $q\bar{q}$ probe are also unaffected by the drag force imparted by the back reacted thermal plasma. We start our analysis with two dimensional uniform motion of a $q \bar{q}$ bound state with a separation length $L$ and the centre of the pair is at the origin of the boundary coordinates.

In the dual gravity theory, we consider a spiral profile of a rotating string with both of its ends are attached at the boundary with the separation length $L$ and the body of the string hanging in to the radial direction $u$ of the bulk described by (\ref{dualgrav}). The ansatz for string profile is given as,
\begin{equation}\label{rgauge1}
X^\mu(\tau,\sigma)=(t=\tau, u=\sigma, x= \rho(\sigma)\cos(\omega t+\theta(\sigma)),
y=\rho(\sigma)\sin(\omega t + \theta(\sigma)), z=0).
\end{equation}
The profile of the string stretches into the bulk up to a certain radial distance implying,
\begin{equation}
\frac{du}{d\rho}=0.
\label{bcqq}
\end{equation}
For pure translational motion of the boundary $q \bar{q}$ probe, the dual string does not experience a drag force and therefore it does not trail behind its endpoints attached to the boundary. However, we are mainly interested in rotational motion of the $q \bar{q}$ probe. Our aim is to holographically show that there should not be any effect of angular drag on the spiral profile of dual rotating string. Using the string profile ansatz (\ref{rgauge1}) we compute the Lagrangian density as
\begin{equation}\label{rlag1}
{\cal L}= [\frac{f^2}{h}(h-\rho^2\omega^2)+f^2(h-\rho^2\omega^2)\rho'^2+hf^2\rho^2\theta'^2]^{\frac{1}{2}},
\end{equation}
The momentum along the direction of $\theta$ and $\rho$ can be derived as,
\begin{eqnarray}
\Pi_\theta^u= \frac{hf^2\rho^2 \theta'}{\cal L} ~~{\rm and}~~
\Pi_\rho^u= \frac{f^2 (h-\omega^2 \rho^2)\rho'}{\cal L}.
\label{momqq}
\end{eqnarray}
By solving (\ref{momqq}) for $\rho'$ and $\theta'$  we get,
\begin{equation}\label{ldas}
\rho' = \frac{\Pi_\rho^u\,\rho}{\sqrt{(h-\omega^2\rho^2)(f^2h\rho^2-\Pi_\theta^{u^2})-h\Pi_\rho^{u^2} \rho^2}},
\end{equation}
and
\begin{equation}\label{thetadas}
\theta' = \frac{\Pi_{\theta}^u (h-\omega^2\rho^2)}{h\rho \sqrt{(h-\omega^2\rho^2)(f^2h\rho^2-\Pi_\theta^{u^2})-h\Pi_\rho^{u^2} \rho^2}}.
\end{equation}
Since at turning point $\rho' $ becomes infinity, the condition for turning point can be achieved by setting the denominator of the right hand side in the equation (\ref{ldas}) to zero.\begin{eqnarray}
(h-\omega^2\rho^2)(f^2h\rho^2-\Pi_\theta^{u^2})\Big|_{u_{\text{turning point}}}  = h \Pi_\rho^{u^2} \rho^2.
\end{eqnarray}
To obtain a non-trivial turning point in this set up we consider $\Pi_\rho^u$ takes non-zero value.
Furthermore, the ratio of  $\rho'$ to $\theta'$ takes the following form,
\begin{equation}
\frac{\rho'}{ \theta'}=\frac{h \rho^2\Pi_\rho^u}{(h-\omega^2\rho^2)\Pi_\theta^u}.
\label{condqq}
\end{equation}
In the left hand side of equation (\ref{condqq}) we get $\rho'=\infty$ at the maxima of the string, so in the right hand side until the condition $h=\omega^2\rho^2$ is met the momentum along the angular direction $\Pi_\theta^u$ should be equal to zero.  In addition to that, to achieve a non trivial value of the turning point we always set $\Pi_\rho^u \ne 0 \Rightarrow h \ne\omega^2\rho^2$. 

Consequently, the $\theta$ equation of motion of the string can be written as,
\begin{equation}
 \partial_t(\Pi_\theta^t) + \partial_u(\Pi_\theta^u) = 0.
\end{equation}
Since the Lagrangian we are interested in is independent from the explicit dependence of time it implies that $\Pi_\theta^u$ is a constant of motion. Therefore, $\Pi_\theta^u$ vanishes not only at the turning point of the string,  but also through out the full string profile. Therefore there is no drag force in the $\theta$ direction and we can conclude that \textit{the rotating $q\bar{q}$ experience no drag in the angular direction.}
It is very interesting to study the translational and rotational motion together and extract the condition for no drag for $q \bar{q}$ bound state. However, the analysis is fully time dependent and requires heavy numerical analysis. We leave this problem for our future study.

\section{Conclusion}\label{con}
In this work, using various holographic methods, we study the effect of back reaction on the hydrodynamical properties of the strongly coupled $\mathcal{N}= 4$ SYM plasma at finite temperature. To estimate the effect of back reaction on the strong coupling properties of the plasma we use heavy quark and $q\bar{q}$ bound state as probes and compute the jet quenching parameter, screening length and binding energy. In each case, we observe that the presence of the back reaction enhances strong coupling effect of the thermal plasma. This observation is consistent with drag force result reported in \cite{Chakrabortty:2011sp}. We also compute the energy loss of a heavy probe quark rotating in the back reacted plasma. We conclude that the presence of back reaction enhances the energy loss.

\text{Gauge/gravity} duality allows us to study these hydrodynamic properties of strongly coupled  back reacted thermal plasma by doing the computation in the corresponding dual gravity theory. In this dual theory, we consider a uniform distribution of infinitely long, static strings hanging from the boundary of the AdS black hole space-time and stretching up to the horizon of the black hole. As a result of introducing this long strings, the AdS black hole gets back reacted and this back reacted geometry is exactly computable. The back reacted  geometry is parameterised by the mass of the black hole as well as long string density. The gravitational stability of the back reacted black hole has been analysed using tensor and vector perturbations.

In this present work, using holographic technic, we study the jet quenching parameter signifying the energy loss due to the suppression of heavy probe quarks with high transverse momentum in the presence of thermal medium. From our analysis, we note that the presence of back reaction always results into enhancement of heavy quark suppression.  We also note that for a fixed value of back reaction the jet quenching parameter monotonously increases with respect to the temperature of the medium.

Moreover we analyse the screening length between the $q\bar{q}$ pair probing the back reacted plasma. It turns out that with the enhancement of back reaction,  the screening length of a $q\bar{q}$ pair reduces significantly. Furthermore, the effect of back reaction also reduces the binding energy of the $q\bar{q}$ pair.

To have a qualitative understanding of the effect of back reaction on energy loss we study the dynamics of a heavy quark rotating with constant angular speed inside the thermal plasma.  Using holographic prescription, we study the ratio of the total energy loss in the presence of back reaction to the total energy loss without back reaction with respect to boundary speed $v=\rho(u\rightarrow 0) \omega$.  When the quark's angular speed is very small the radiation energy loss naturally remains insignificant. However, it is evident from the plot that the effect of back reaction significantly enhances the total energy loss. This particular observation leads to the conclusion that the presence of back reaction actually results in to the enhancement of the drag energy loss. Moreover, as the linear speed of quark approaches to the speed of light, it overcomes the drag force imparted by the back reacted plasma.  When the quark's angular speed increases sufficiently the ratio takes values very close to unity and this observation implies that the dominating contribution for total energy loss comes from the radiation effect. The study of the ratio of the total energy loss to the drag energy loss gives more support to these conclusions. We also plot the ratio of total energy loss in the back reacted thermal plasma to the energy loss in the vacuum of the theory. The plot clearly shows that for a lower angular speed of probe quark, the ratio takes very high values. This signifies that the energy loss is fully dominated by the drag effect. For a higher angular speed, the ratio becomes close to unity and it implies that radiation effect contributes more to the total energy loss of the probe quark.

Finally we have studied the dynamics of a heavy rotating $q \bar{q}$ pair in the back reacted thermal plasma.  We show that in the case of pure rotational motion the dynamics of the $q \bar{q}$ probe is free of angular drag.

It is important to note that in the phenomenological study of hydrodynamical aspects associated with the QGP medium, the back reaction of the plasma is usually neglected. The back reaction we consider can be created by adding other heavy static quarks in the thermal plasma. Within the regime of \textit{gauge/gravity} duality, our present work perhaps makes an effort to capture such back reaction effect produced by those other heavy quarks present in the plasma.

 \section*{Acknowledgements}
The authors would like to acknowledge Gokhan Alkac, Arjun Bagchi, Rudranil Basu, Eric A. Bergshoeff, Sudipta Mukherji, Sunil Mukhi, V.A. Penas, Shibaji Roy, Bala Sathiapalan, Hesam Soltanpanahi for various fruitful discussion. SC is supported by Erasmus Mundus NAMASTE India-EU Grants. SC thanks the Indian Institute of Science Education and Research-Pune and the Institute of Mathematical Sciences for the hospitality during a major part of this work. TD thanks Institute of Physics for the hospitality during the initial part of this work.
\vspace{.5cm}
\bigskip

\

\end{document}